\documentclass[aps,pra,twocolumn,superscriptaddress,groupedaddress,amsfonts,amsmath,floatfix]{revtex4}

\usepackage{graphicx}
\usepackage{color}
\usepackage{subfigure}
\usepackage{bm}
\usepackage{epstopdf}

\begin{document}
\title{Vortices in spin-orbit-coupled Bose-Einstein condensates}

\author{J. Radi\'{c}}
\affiliation{Joint Quantum Institute, University of Maryland, College Park, Maryland
20742-4111, USA}
\author{T. A. Sedrakyan}
\affiliation{Joint Quantum Institute, University of Maryland, College Park, Maryland
20742-4111, USA}
\author{I. B. Spielman}
\affiliation{Joint Quantum Institute, University of Maryland, College Park, Maryland
20742-4111, USA}
\affiliation{National Institute of Standards and Technology, Gaithersburg, Maryland
20899, USA}
\author{V. Galitski}
\affiliation{Joint Quantum Institute, University of Maryland, College Park, Maryland
20742-4111, USA}

\date{\today}

\begin{abstract}
Realistic methods to create vortices in spin-orbit-coupled Bose-Einstein condensates are discussed. It is shown that, contrary to common intuition, rotation of the trap containing a spin-orbit condensate does not lead to an equilibrium state with static vortex structures, but gives rise instead to non-equilibrium behavior described by an intrinsically time-dependent Hamiltonian. We propose here the following alternative methods to induce thermodynamically stable static vortex configurations: (1)~to rotate both the lasers and the anisotropic trap; and (2)~to impose a synthetic Abelian field on top of synthetic spin-orbit interactions. Effective Hamiltonians for spin-orbit condensates under such perturbations are derived for most currently known realistic laser schemes that induce synthetic spin-orbit couplings. The Gross-Pitaevskii equation is solved for several experimentally relevant regimes. The new interesting effects include spatial separation of left- and right-moving spin-orbit condensates, the appearance of unusual vortex arrangements, and parity effects in vortex nucleation where the topological excitations are predicted to appear in pairs. All these phenomena are shown to be highly non-universal and depend strongly on a specific laser scheme and system parameters. 
\end{abstract}


\maketitle

\section{Introduction}

Spin-orbit-coupled cold atoms represent a very new and quickly growing area of research that promises to host an even richer variety of exotic phenomena than solid-state spintronics \cite{Zutic2004}. Indeed, within just a few years of experimental research in the field, a number of exciting phenomena have already been observed \cite{Lin2009_2,Lin2009,Lin2011,Lin_electric} and there are clearly many more low-hanging fruits awaiting their experimental discovery.

The key ideas underlying cold-atom spintronics - that studies particles with a synthetic spin degree of freedom coupled to their motion - grew out of the early theoretical work by Juzeli$\bar{\mbox{u}}$nas et al. \cite{Juzeliunas2004,Juzeliunas2005,Ruseckas2005,Jaksch2003,Osterloh2005,Zhu2006,Satija2006}, 
which showed that single-particle physics of atom-laser dressed states, where internal atomic states are coupled by position-dependent laser fields, can be described in terms of a non-Abelian vector potential acting on the dressed excitations. Later, it was demonstrated theoretically \cite{Stanescu2007} that specific realizations of such laser configurations, including the early-proposed tripod scheme, give rise to spin-orbit-coupled Hamiltonians of Rashba-Dresselhaus type, familiar from solid-state semiconductor spintronics and that this ``spintronics'' description is a convenient alternative to the description in terms of the non-Abelian fields. Most importantly, it was quickly realized \cite{Stanescu2008} that contrary to solid-state spintronics, where the underlying particles are bound to be electronic excitations, the synthetic spin-1/2 degree of freedom in cold atoms can be carried by dressed spin-orbit-coupled bosons that were predicted  to condense into a state, dubbed in Ref. \cite{Stanescu2008} a ``spin-orbit coupled Bose-Einstein Condensate (BEC).'' It was also shown~\cite{Stanescu2008} that multiple peaks in the time-of-flight expansion would be a smoking gun signature of such a new quantum state. Remarkably, this type of behavior was observed experimentally~\cite{Lin2009_2} by one of the authors shortly after. The specific laser setup used in Ref.~[\onlinecite{Lin2009_2}] - that gives rise to an ``Abelian'' spin-orbit coupling (sometimes referred to as the ``persistent-spin-helix symmetry point,''~\cite{Schliemann2003,Chalaev2005,Bernevig2006,Stanescu2007:PRB,Duckheim2009,Koralek2009,Hasan2010} where the Rashba and Dresselhaus couplings are equal to each other) - was later analyzed in detail by  Ho and collaborators in Ref.~[\onlinecite{Ho2010}]. These experimental and theoretical successes have motivated other interesting theoretical proposals for realistic experimental schemes that can be used to create spin-orbit-coupled systems~\cite{Juzeliunas2010,Campbell2011}. Spin-orbit-coupled BECs have also been studied theoretically in Refs.~[\onlinecite{Larson2009,Wang2010,Yip2011,Zhang2011,Xu2011,Kawakami2011,Jian2011,DW_Zhang2011,Gopalakrishnan2011,Wu2011,Dalibard2010}]
for different types of spin-orbit interactions and different  internal structures of bosons (pseudospin-1/2, spin-1 and spin-2 bosons).

Among the obvious questions about the spin-orbit BECs is the physics of topological excitations - vortices - that play a central role in the physics of conventional BECs. This is subject of this paper, where we focus primarily on exploring experimentally-relevant methods that can be used to nucleate static vortex structures in spin-orbit BECs. In contrast to the conventional condensates, the situation here is shown to be significantly more complicated as the vortex physics is obscured by the interplay of external perturbations intended to create them and the hyperfine structure underlying the synthetic spin-orbit-coupling setup.

It is widely known and often taken for granted that rotating a Bose-Einstein condensate gives rise to
the formation of vortices that arrange themselves into static vortex lattice structures. However, this picture is not in
fact an obvious outcome of rotation, which represents a {\em time-dependent perturbation} due to a rotating anisotropic trap potential. The many subtleties involved in understanding the fundamentals of the related phenomena are discussed in detail in the reviews by Leggett \cite{Leggett2001,
Leggett2000}, but the main conclusion is indeed that the physics of a one-component BEC confined to a spinning anisotropic trap can be mapped onto a statistical-mechanical  problem of the BEC with an effective {\em time-independent} Hamiltonian, $H_{\rm eff} = H -\bm{\omega}_r \cdot {\bf L}$, which describes the system in a rotating frame of reference (here, ${\bf L}$ is the orbital angular momentum operator and $\bm{\omega}_r$ is the frequency of rotation).

A na{\"\i}ve expectation therefore is that to rotate an anisotropic trap would be a straightforward means to create vortex structures in spin-orbit-coupled BECs as well. However, this paper shows that this is  generally {\em not so} and other, more sophisticated methods have to be involved in order to create static vortex structures. We show that the problem with rotation arises here because atoms are not influenced by the trapping potential  only, but also by the lasers which
create spin-orbit coupling in the first place. Therefore, if only the anisotropic potential rotates, it is in general impossible to choose a frame of reference where the Hamiltonian is time-independent, because the ``spin-orbit coupling'' lasers, stationary in the lab frame, are rotating in the rotating frame, generally resulting in non-trivial dynamics in any rotating frame. While there do exist rare degenerate cases, where a unitary transformation that eliminates time-dependence from the non-interacting Hamiltonian can be explicitly found,  the interaction terms generally become time-dependent under the unitary transformation, resulting again in a non-equilibrium problem.  Hence, we argue that the residual time-dependence appears to be an essential and unwelcome property of a spin-orbit-coupled BEC with a rotating anisotropic potential (at least for the realistic laser schemes currently known to us). We believe that while the specifics of time-evolution of rotating spin-orbit BECs are sensitive to details of both the laser setup used and interactions, the typical scenario will involve non-universal dynamics that would inevitably lead to heating and destruction of the coherent state in contrast to the conventional BECs.

It is therefore desirable to develop other experimentally-relevant methods to create vortices, like rotation or a magnetic field, for spin-orbit-coupled BECs. Two other ways suggested here and examined in detail are as follows: (i)~to rotate both the lasers creating spin-orbit coupling and the trap, if the latter is anisotropic, or just the lasers for an isotropic trap (note that to rotate an isotropic trap has no meaning); (ii)~To combine synthetic spin-orbit-couplings with a synthetic Abelian magnetic field. Theoretically, both methods are shown to give rise to interesting phenomena, including the appearance of sought-after static vortices and vortex lattices, parity effects in vortex nucleation, and real-space splitting of the spin-orbit BEC where the left- and right-moving parts are physically separated (an effect, which bears some similarity to the spin-Hall effect~\cite{SHE_theor,SHE_exp} known in condensed matter spintronics). 

Our paper is structured as follows: Sec.~\ref{Sec:different_schemes} derives effective  Hamiltonians
corresponding to a rotating trapping potential and/or rotating ``spin-orbit lasers'' for various spin-orbit-coupled laser schemes. In Sec.~\ref{Sec:vortices_rotation}, we solve the Gross-Pitaevskii equation to describe individual vortices and collective vortex structures for the laser scheme described in Ref.~[\onlinecite{Lin2011}] with a rotating trap and Raman lasers. In Sec.\ref{Sec:detuning}, we investigate vortex nucleation and other effects associated with a synthetic magnetic field that can be imposed on top of the spin-orbit coupled system used in \cite{Lin2011} by applying a spatially dependent Zeeman field.

\section{Rotation in systems with engineered spin-orbit coupling}
\label{Sec:different_schemes}

In this section, we investigate the effect of rotation of an anisotropic trapping
potential and/or spin-orbit lasers in three different laser schemes that have been proposed to 
create effective spin-orbit couplings. To distinguish between the different schemes, we will refer to the
 setup used in Ref.~[\onlinecite{Lin2011}] as ``M-scheme,'' the proposal described in Refs.~[\onlinecite{Ruseckas2005,Stanescu2007}] as ``tripod-scheme,'' and the recent proposal of Ref.~[\onlinecite{Campbell2011}]
 as ``4-level-scheme.''

\subsection{M-scheme}
\label{Subsec:Nature_scheme}

  We first focus on the scheme used in recent experiment \cite{Lin2011} and investigate the Hamiltonian
for the case in which both trap and spin-orbit coupling lasers are rotating about the z-axis.
The atoms in \cite{Lin2011} are under the influence of three external sources: trapping potential, 
Raman lasers which create spin-orbit coupling and magnetic field which creates Zeeman splitting (aligned
along y-direction).
If we wanted to get a time-independent Hamiltonian in the rotating frame we would have to rotate
trapping potential, Raman lasers and magnetic field. To make things easier it is possible to change direction
of the magnetic field to be along z-axis, which makes rotation of magnetic field about the z-axis unnecessary.
If the change of the direction of magnetic
field is accompanied by change in polarization of Raman lasers (the direction of lasers stays the same)
the system is described by the same effective equations as in \cite{Lin2011}.   
  It is also important to note that, in the case of an isotropic trap, rotation of the trap
does not have any effect and in that case rotating only the Raman lasers suffices.
  The stationary system is described by the following 
Hamiltonian (see methods in \cite{Lin2011}):
\begin{equation}
\begin{split}
\hat{H}_0 & = \left[ \frac{\hbar^2 \hat{{\mathbf k}}^2}{2m} + V({\mathbf r})  \right] \check{1} +
\begin{pmatrix}
\hbar \left( -\omega_z + \omega_q \right)& 0& 0 \\
0& 0& 0 \\
0& 0& {\hbar \omega_z} 
\end{pmatrix} \\ 
& \quad + \sqrt{2}\Omega \check{\sigma}_{3,x} \cos(2 k_L x + \Delta \omega_L t),
\end{split}
\label{eq:basic_Hamiltonian}
\end{equation}
where $\hat{{\mathbf k}}=-i\bm {\nabla}$, $V({\mathbf r})$ is the trapping potential, 
$\check{1}$ is the 3 $\times$ 3 identity matrix, $\check{\sigma}_{3,x,y,z}$ are the 3 $\times$ 3 
spin matrices,  $k_L=\sqrt{2} \pi/\lambda $, $\Omega$ is the
Raman coupling strength, $\omega_z$ and
$\omega_q$ are the linear and quadratic Zeeman shifts, respectively. Here $\lambda$ is the wavelength 
and $\Delta \omega_L$ is the frequency difference of the two Raman beams used in the M-scheme.  
The Hamiltonian is written in the basis of hyperfine states
$\lbrace \lvert m_F=+1 \rangle, \lvert m_F=0 \rangle, \lvert m_F=-1 \rangle \rbrace$ which are
quantized in $\hat{z}$ direction (direction of the external magnetic field). 

When the trap and Raman lasers rotate with a constant frequency $\omega_{r}$ about the z-axis, the Hamiltonian 
$\hat{H}_{\rm rot}$ in the laboratory frame can be obtained from Eq.~(\ref{eq:basic_Hamiltonian}) using the following 
substitutions:
\begin{equation*}
V(x,y,z) \rightarrow V(x(t),y(t),z)
\end{equation*}
\begin{equation}
\check{\sigma}_{3,x} \cos(2 k_L x + \Delta \omega_L t)
\rightarrow \check{\sigma}_{3,x}(t) \cos(2 k_L x(t) + \Delta \omega_L t),
\end{equation}
where
\begin{equation}
\begin{split}
& \quad x(t)=x \cos(\omega_r t) + y \sin(\omega_r t) \\
& \quad y(t)=y \cos(\omega_r t) - x \sin(\omega_r t) \\
& \check{\sigma}_{3,x}(t)=\check{\sigma}_{3,x} \cos(\omega_r t) + \check{\sigma}_{3,y} \sin(\omega_r t).
\end{split}
\label{eq:xy_t}
\end{equation}
The Hamiltonian $\hat{H}_{\rm rot}$ can be also written in a more compact form:
\begin{equation}
\hat{H}_{\rm rot}=e^{-i \omega_r t (\hat{L}_z+\hat{S}_z)/\hbar} \hat{H}_0 e^{i \omega_r t (\hat{L}_z+\hat{S}_z)/\hbar},
\label{eq:H_rot}
\end{equation}
where $\hat{\bf L}$ is the orbital angular momentum operator and $\hat{\bf S}$ is the spin operator, and $\hat{L}_z$ and $\hat{S}_z$ are their $z$-components: $\hat{L}_z=\hbar \big(x \hat{k}_y - y \hat{k}_x \big) \check{1}$, $\hat{S}_z=\hbar \check{\sigma}_{3,z}$. 

The Hamiltonian (\ref{eq:H_rot}) is time-dependent in the laboratory frame, but we show below that this time-dependence can be eliminated by a unitary transform. Recall that an arbitrary unitary transform, $\hat{U}(t)$, of the Hamiltonian $\hat{H}$ produces a new Hamiltonian, $\hat{H}^{\prime}$, as follows
\begin{equation}
\hat{H}^{\prime}=\hat{U} \hat{H} \hat{U}^{\dagger} - i \hbar \hat{U} \frac{\partial \hat{U}^{\dagger}}{\partial t}.
\label{eq:general_transformation}
\end{equation}

 We first go to the rotating frame of reference (rotating together with both the trap and the lasers) 
\cite{Castin1999}: $\lvert \psi_{RF} \rangle=\hat{U}(t) \lvert \psi \rangle$, where $\hat{U}(t)=\exp [i\omega_r t (\hat{L}_z + \hat{S}_z)/\hbar]$.  Eq.~(\ref{eq:general_transformation}) yields
\begin{equation}
\hat{H}_{RF}=\hat{H}_0-\omega_r (\hat{L}_z + \hat{S}_z),
\label{eq:H_RF}
\end{equation}
where $\hat{H}_{RF}$ denotes the Hamiltonian in the rotating frame. The remaining time-dependence, arising from the oscillating Raman laser fields in $\hat{H}_0$, can be removed in the framework of the rotating wave approximation.
To obtain an effective description of the system in terms of  two internal pseudo-spin states, we follow \cite{Lin2011} and choose the quadratic Zeeman shift  $\hbar \omega_q$ to be large  enough, so that the state $\lvert m_z=1 \rangle$ can be neglected. Using the pseudo-spin-1/2 labels for internal states, we get,
$\lvert \uparrow \rangle \equiv \lvert m_z=0 \rangle$, $\lvert \downarrow \rangle \equiv \lvert m_z=-1 \rangle$. 
The final Hamiltonian can be expressed in the form used in Ref.~[\onlinecite{Lin2011}]  (a detailed derivation is much analogous to  Ref.~[\onlinecite{Lin2011}]  and is presented in appendix~\ref{app:scheme1_a}) as follows,
\begin{equation}
\begin{split}
& \hat{H}_{RF,2} =\left[ \frac{\hbar^2 \hat{{\mathbf k}}^2}{2m} + V({\mathbf r}) 
- \omega_{r} \hat{L}_z +E_L \right] \check{1} \\
& \quad + \frac{\hbar^2 k_L}{m} \hat{k}_x \check{\sigma}_z
+ \frac{\Omega}{2} \check{\sigma}_x
+ \hbar \omega_{r} k_L y \check{\sigma}_z 
+ 
\begin{pmatrix}
0& 0\\
0& \hbar \omega_r - \delta
\end{pmatrix},
\end{split}
\label{eq:Nature_final}
\end{equation}
where $\check{1}$ is 2 $\times$ 2 unit matrix, $\check{\sigma}_{x,y,z}$ are 2 $\times$ 2
Pauli matrices and $\delta=\hbar(\Delta \omega_L - \omega_z)$ is a detuning from the Raman resonance.
Since the resulting Hamiltonian is time-independent in the rotating frame, it leads to the appearance of stationary vortex structures studied below in Sec.~\ref{Sec:vortices_rotation}. 

 In the case where {\em only} the anisotropic trap is rotating, the Hamiltonian in the laboratory frame is given by (\ref{eq:basic_Hamiltonian}), with $V \big(x,y,z \big) \rightarrow V \big( x(t),y(t),z \big)$. 
Importantly, if we go to the rotating frame and make the rotating wave approximation (exactly as in the 
above), we are still left with a time-dependence (for details see appendix \ref{app:scheme1_b}):
\begin{equation}
\begin{split}
\hat{H}_{RF,2}^{\prime}=& \left[ \frac{\hbar^2 \hat{{\mathbf k}}^2}{2m} + V({\mathbf r}) 
- \omega_{r} \hat{L}_z \right] \check{1} \\
& +\frac{\hbar^2 k_L }{m} \hat{k}_x(t) \check{\sigma}_z
+ \frac{\Omega}{2} \check{\sigma}_x
+ \frac{\delta}{2} \check{\sigma}_z,
\end{split}
\label{eq:Nature_trap_equation}
\end{equation}
where $\hat{k}_x(t)=\hat{k}_x \cos(\omega_r t) - \hat{k}_y \sin(\omega_r t)$.

\subsection{Tripod scheme}
\label{Subecs:Tripod_scheme}

  We now concentrate on the proposal described in Refs.~[\onlinecite{Ruseckas2005,Stanescu2007}], which uses a 
  so-called ``tripod scheme,'' that consists of three degenerate ground states of an atom  coupled  to an excited state. The resulting energy spectrum includes two degenerate ``dark'' states and two ``bright''
states (one of the bright states is higher and the other is lower in energy with respect to 
degenerate dark states). In the strong coupling regime and within the adiabatic approximation, the energy difference between the dark and bright states is very large compared to other characteristic energies of the system. In this case, a coupling between the dark and bright
states is negligible, and consequently if the atoms initially exist within the dark states subspace, they are expected 
to stay there for a long time. From now on, we use pseudo-spin-1/2 notations for the two degenerate dark states.

 The effective stationary Hamiltonian (projected onto the dark-state subspace) reads:
\begin{equation}
H=\left[ \frac{\hat{{\mathbf p}}^2}{2m} + w({\mathbf r}) \right] \check{1} - v_0 \hat{p}_x \check{\sigma}_y 
- v_1 \hat{p}_y \check{\sigma}_z + \delta_0 \check{\sigma}_z,
\label{eq:tripod_stationary}
\end{equation}
where ${\mathbf p}=-i \hbar \bm{\nabla}$, $w({\mathbf r})$ is a spin-independent part of the trapping potential (see appendix~ \ref{app:scheme_2} for details), $v_0$ and $v_1$ characterize the strength and type of spin-orbit 
coupling, and $\delta_0$ is the effective Zeeman splitting. $\check{1}$ is a 2 $\times$ 2 unit matrix, $\check{\sigma}_{x,y,z}$ are 2 $\times$ 2 Pauli matrices.

We first investigate the case with  both the trap and the spin-orbit lasers rotating.
The derivation, presented in appendix~\ref{app:scheme_2}, leads to the following Hamiltonian in the rotating frame
\begin{equation}
\begin{split}
&\hat{H}_{RF,2}=\left[ \frac{\hat{{\mathbf p}}^2}{2m} + w({\mathbf r}) - \hat{L}_z \right]\check{1} 
-v_0 \hat{p}_x \check{\sigma}_y - v_1 \hat{p}_y \check{\sigma}_z  \\
& \quad + \delta_0 \check{\sigma}_z + m \hbar \omega_r (v_1 x \check{\sigma}_z - v_0 y \check{\sigma}_y) \\
& \quad - \hbar \omega_r
\begin{pmatrix}
\sin^2 \phi & \sin \phi \cos \phi \cos \theta \\
\sin \phi \cos \phi \cos \theta & \cos^2 \theta \cos^2 \phi - \sin^2 \theta
\end{pmatrix},
\end{split}
\label{eq:tripod_final}
\end{equation}
where $\phi=m v_0 x/ \cos \theta$, $\delta_0=\sin^2 \theta \big \lbrace \delta - 
\big[ \big( \frac{v_0}{\cos \theta} \big)^2 + \big( \frac{v_1}{\sin^2 (\theta/2)} \big)^2 \big]/2 
\big \rbrace /2$, and $\theta$ is a constant. Let us note here that Ref.~[\onlinecite{Burrello2010}] 
previously considered the tripod scheme under rotation, but obtained slightly different results
(the spin angular momentum part ($-\omega_r \hat{S}_z$) was ignored in Ref.~[\onlinecite{Burrello2010}] ). 

Our result (\ref{eq:tripod_final}), together with Eq.~(\ref{eq:Nature_final}) for the M-scheme, clearly shows
that the effect of rotation in systems with synthetic spin-orbit interaction does not  reduce to just adding the $-\omega_r L_z$ term for the Hamiltonian in the rotating frame, but also produces other position-dependent terms, which depend on a particular scheme.

  We now consider the tripod scheme with only the trap rotating. 
We first address the following question: if the trapping potential is time-dependent, 
can we get the effective pseudo-spin Hamiltonian in the laboratory frame just by changing $V \rightarrow V(t)$ in (\ref{eq:tripod_stationary}); or in other words, are we still  allowed to restrict to the dark-state subspace if the external potential is time dependent? 
The answer is certainly ``yes,'' if the trapping potential is the same for all three degenerate ground states
(which is most often the case for optical trapping), because this kind of time-dependent potential does not 
couple the dark and bright states. 

In a general tripod scheme however, the trapping potential is not spin-independent ($\hat{V}({\mathbf r})=\sum_j V_j({\mathbf r})\lvert j \rangle \langle j \rvert$, $V_1=V_2=w({\mathbf r})$ and $V_3=
w({\mathbf r})+\delta$). To better understand this case, let us choose states $\lbrace \lvert 1 \rangle, \lvert 2 \rangle, \lvert 3 \rangle \rbrace$ to be eigenstates of $\hat{S}_z$ (z-component of the total spin operator). Then, the rotation of the trapping potential about the z-axis is described by:
$V_1=V_2=w^{\prime}({\mathbf r},t)$ and $V_3=w^{\prime}({\mathbf r},t)+\delta$, where $w^{\prime}({\mathbf r},t)=
e^{-i\omega_r t \hat{L}_z/\hbar}w({\mathbf r})e^{i\omega_r t \hat{L}_z/\hbar}$. We can therefore separate $\hat{V}({\mathbf r})$ into a stationary spin-dependent term and  a time-dependent but spin-independent term:
$\hat{V}({\mathbf r},t)=\delta \lvert 3 \rangle \langle 3 \rvert + w^{\prime}({\mathbf r},t) \big( \lvert 1 \rangle \langle 1 \rvert +\lvert 2 \rangle \langle 2 \rvert +\lvert 3 \rangle \langle 3 \rvert \big)$. Therefore the time-dependent
part of trapping potential is spin-independent and it will not couple dark and bright states.
With this, the tripod system with a trap rotating about the z-axis is described by
\begin{equation}
\hat{H}=\left[ \frac{\hat{{\mathbf p}}^2}{2m} + w^{\prime}({\mathbf r},t) \right] \check{1} 
- v_0 \hat{p}_x \check{\sigma}_y - v_1 \hat{p}_y \check{\sigma}_z + \delta_0 \check{\sigma}_z.
\label{tripod_time}
\end{equation}
We now make the following transformation.
$\hat{U}(t)=\exp[i \omega_r t (\hat{L}_z/\hbar 
+ \check{\sigma}_x/2)]$, which gives:
\begin{equation}
\begin{split}
\hat{H}^{\prime} &=\left[ \frac{\hat{{\mathbf p}}^2}{2m} + w({\mathbf r}) - \omega_r \hat{L}_z \right] \check{1} 
- v_0 \hat{p}_x(t) \check{\sigma}_y(t) \\
&- v_1 \hat{p}_y(t) \check{\sigma}_z(t) + \delta_0 \check{\sigma}_z(t) - \frac{\hbar \omega_r}{2} \check{\sigma}_x,
\end{split}
\label{eq:tripod_trap_u}
\end{equation}
where
\begin{equation}
\begin{split}
& \quad \hat{p}_x(t)=\hat{p}_x \cos(\omega_r t) - \hat{p}_y \sin(\omega_r t), \\
& \quad \hat{p}_y(t)=\hat{p}_y \cos(\omega_r t) + \hat{p}_x \sin(\omega_r t), \\
& \check{\sigma}_{y}(t)=\check{\sigma}_{y} \cos(\omega_r t) - \check{\sigma}_{z} \sin(\omega_r t), \\
& \check{\sigma}_{z}(t)=\check{\sigma}_{z} \cos(\omega_r t) + \check{\sigma}_{y} \sin(\omega_r t).
\end{split}
\label{eq:px_py_t}
\end{equation}
The Hamiltonian (\ref{eq:tripod_trap_u}) is generally time-dependent.
However in the case of Rashba coupling ($v_0=v_1=v$) and $\delta_0=0$,
this  {\em non-interacting part} of the Hamiltonian becomes static and reads,
\begin{equation}
\begin{split}
\hat{H}^{\prime} & =\left[ \frac{\hat{{\mathbf p}}^2}{2m} + w({\mathbf r}) - \omega_r \hat{L}_z \right] \check{1} \\
& \quad - v \left( \hat{p}_x \check{\sigma}_y + \hat{p}_y \check{\sigma}_z \right)
- \frac{\hbar \omega_r}{2} \check{\sigma}_x.
\end{split}
\label{tripod_frame}
\end{equation}

\subsection{4-level scheme}
\label{Subsec:4-pod}

 Here we study the 4-level-scheme~\cite{Campbell2011} with a rotating trap.
The stationary effective Hamiltonian (projected to the lowest energy states) is given by~\cite{Campbell2011}:
\begin{equation}
\begin{split}
\hat{H}=&\left[ \frac{\hbar^2 \hat{{\mathbf k}}^2}{2m} + V({\mathbf r}) \right] \check{1} 
+ \alpha(\check{\sigma}_x \hat{k}_y - \check{\sigma}_y \hat{k}_x) \\
&+ \beta(\check{\sigma}_x \hat{k}_y + \check{\sigma}_y \hat{k}_x)
+\frac{\Delta_z}{2} \check{\sigma}_z,
\end{split}
\label{4-pod}
\end{equation}
where $\alpha$ and $\beta$ denote strengths of Rashba and Dresselhaus couplings respectively (in this scheme, $\alpha$  is fixed and $\beta$ can be tuned), and $\Delta_z$ is an effective Zeeman field.
Per the same arguments as in the tripod scheme, we are allowed to simply replace $V \rightarrow V(t)$ in
(\ref{4-pod}) (if an external potential is time-dependent; note also, that the trapping potential here is spin independent). 
The rotating trap potential reads: $V({\mathbf r},t)=e^{-i\omega_r t \hat{L}_z/\hbar}V({\mathbf r})e^{i\omega_r t \hat{L}_z/\hbar}$. We now make the following transformation: $\hat{U}(t)=\exp[i \omega_r t (\hat{L}_z/\hbar + \check{\sigma}_z/2)]$, which gives:
\begin{equation}
\begin{split}
\hat{H}^{\prime} & =\left[ \frac{\hbar^2 \hat{{\mathbf k}}^2}{2m} + V({\mathbf r}) - \omega_r \hat{L}_z \right] \check{1} 
+ \alpha (\check{\sigma}_x \hat{k}_y - \check{\sigma}_y \hat{k}_x) \\
&+ \beta \bigg \lbrace \big[ \check{\sigma}_y \cos(2 \omega_r t) + \check{\sigma}_x \sin(2 \omega_r t) \big] \hat{k}_x\\
&+ \big[ \check{\sigma}_x \cos(2 \omega_r t) - \check{\sigma}_y \sin(2 \omega_r t) \big] \hat{k}_y \bigg \rbrace \\
&+ \left( \frac{\Delta_z}{2} - \frac{\hbar \omega_r}{2} \right) \check{\sigma}_z.
\end{split}
\label{4-pod_time}
\end{equation}
Again, this {\em non-interacting part} of the Hamiltonian is in general time-dependent, however for pure Rashba coupling, it becomes time-independent.

Note that to get the full Hamiltonian in the rotating frame, we must also include interactions between the bosons and apply to them the same transformations as in the non-interacting part above. If both the trap and spin-orbit lasers rotate, the corresponding unitary operator, $\hat{U}(t)=\exp [i\omega_r t (\hat{L}_z + \hat{S}_z)/\hbar]$, describes a spatial rotation about the z-axis. If the bare interactions are rotationally-invariant, the interaction part of the Hamiltonian does not change in the rotating frame. In contrast to this result however, if only the trap is rotating, the interactions  will generally acquire time-dependence as well (we have found a few very special cases - with serious constraints on the parameters of the system - where a unitary transform can be found that makes both the pure Rashba non-interacting part and interactions time-independent, but whether these degenerate cases can be realized experimentally remains unclear at this stage).


\section{Creating vortices by rotation}
\label{Sec:vortices_rotation}

  In the previous section, we have shown that the Hamiltonian for the M-scheme 
in the presence of a rotating trap and Raman lasers becomes time-independent in the 
rotating frame. In analogy with the physics of ``ordinary'' BEC under rotation, there will be thermal equilibration in the system and vortices will form in the condensate.
     

Let us assume that $\hbar \omega_z \gg \mu$ ($\mu$ is the chemical
potential), which gives an effective 2D system, where the motion in z-direction is effectively frozen (this can
be achieved by applying a 1D optical lattice in $\hat{z}$ direction).  
We also assume the  interaction part of the Hamiltonian to have the form:
\begin{equation}
\hat{H}_{int}=\int d^2r \left( \frac{1}{2} G_1 \hat{\rho}_{\uparrow}^2
+ \frac{1}{2} G_2 \hat{\rho}_{\downarrow}^2 
+ G_{12} \hat{\rho}_{\uparrow} \hat{\rho}_{\downarrow} \right),
\label{H_int}
\end{equation}
where $G_1$, $G_2$ and $G_{12}$ are effective 2D interaction strengths and are related to 3D interaction strengths: $G_1=G_{1}^{3d}/(\sqrt{2\pi} l_z)$, $G_2=G_{2}^{3d}/(\sqrt{2\pi} l_z)$ and $G_{12}=G_{12}^{3d}/(\sqrt{2\pi} l_z)$, where
$l_z=\sqrt{\hbar/(m \omega_z)}$. $\hat{\rho}_{\uparrow}$ and $\hat{\rho}_{\downarrow}$ are density operators for
$\lvert \uparrow \rangle$, $\lvert \downarrow \rangle$ states (normal ordering of the corresponding 
creation/annihilation operators is implied). 
  
  We are interested in finding the ground state configuration of bosons
in a rotating system described by (\ref{eq:Nature_final},\ref{H_int}). First, we have to make
an assumption about the ground state and we assume below that (at the mean-field level) all
 atoms occupy the same single-particle state described by the spinor wave-function, $\big( \psi_{\uparrow}({\mathbf r}), 
\psi_{\downarrow}({\mathbf r}) \big)$ (we also call it condensate wave-function).
The condensate wave-function  satisfies the Gross-Pitaevskii (GP) equations below:
\begin{multline*}
\mu \psi_{\uparrow}=
\bigg[-\frac{\hbar^2}{2m} \nabla^2 - i \frac{\hbar^2 k_L}{m} \frac{\partial}{\partial x}
+ V({\mathbf r}) 
-\omega_{r} \big( \hat{L}_z  - \hbar k_L y \big) \\
+ N G_1 |\psi_{\uparrow}|^2 + N G_{12} |\psi_{\downarrow}|^2
\bigg] \psi_{\uparrow} + \frac{\Omega}{2} \psi_{\downarrow}
\end{multline*}
\begin{multline}
\mu \psi_{\downarrow}=
\bigg[-\frac{\hbar^2}{2m} \nabla^2 + i \frac{\hbar^2 k_L}{m} \frac{\partial}{\partial x}
+ V({\mathbf r}) 
-\omega_{r} \big( \hat{L}_z + \hbar k_L y - \hbar \big) \\
-\delta + N G_2 |\psi_{\downarrow}|^2 + N G_{12} |\psi_{\uparrow}|^2
\bigg] \psi_{\downarrow} + \frac{\Omega}{2} \psi_{\uparrow}
\label{GP_time_independent}
\end{multline}
where $N$ is the total number of particles and $\mu$ is the Lagrange multiplier 
associated with the constraint $\int d^2r \big( |\psi_{\uparrow}|^2 + 
|\psi_{\downarrow}|^2 \big)=1$ (it can be shown that $\mu$ has a physical meaning of chemical potential \cite{Leggett2001}).
We solve the GP equations by using norm-preserving imaginary time propagation
method (see for example Ref.~\cite{Dalfovo1996,Castin1999}). 

  We consider a trapping potential of the following form: $V=\frac{1}{2}m\omega^2
\big( x^2 + \gamma^2 y^2 \big)$, where $\omega$ and $\gamma \omega$ are trapping frequencies
in the $\hat{x}$ and $\hat{y}$ direction. It is convenient to measure lengths in the units of  the 
harmonic oscillator length, $a_0=\sqrt{\hbar/(m \omega)}$ and energy in terms of
 $\hbar \omega$. We introduce dimensionless position variable
${\mathbf r}^{\prime}={\mathbf r}/a_0$.  The corresponding ``dimensionless GP equations'' reads
\begin{multline*}
\mu \psi_{\uparrow}=
\bigg[-\frac{1}{2} \nabla^{\prime 2} - i k_L^{\prime} \frac{\partial}{\partial x^{\prime}}
+ \frac{1}{2} \big( x^{\prime 2} + \gamma^2 y^{\prime 2} \big) \\
- \omega_{r}^{\prime} \big( \hat{L}^{\prime}_z - k_L y^{\prime} \big)
+ g_1 |\psi_{\uparrow}|^2 + g_{12} |\psi_{\downarrow}|^2
\bigg] \psi_{\uparrow} + \frac{\Omega^{\prime}}{2} \psi_{\downarrow}
\end{multline*}
\begin{multline}
\mu \psi_{\downarrow}=
\bigg[-\frac{1}{2} \nabla^{\prime\ 2} + i k_L^{\prime} \frac{\partial}{\partial x^{\prime}}
+ \frac{1}{2} \big( x^{\prime 2} + \gamma^2 y^{\prime 2} \big) -\delta^{\prime}\\
-\omega_{r}^{\prime} \big( \hat{L}^{\prime}_z + k_L y^{\prime} - 1 \big) 
+ g_2 |\psi_{\downarrow}|^2 + g_{12} |\psi_{\uparrow}|^2
\bigg] \psi_{\downarrow} + \frac{\Omega^{\prime}}{2} \psi_{\uparrow},
\label{dimensionless_GP}
\end{multline}
where $k_L^{\prime}=k_L a_0$, $\Omega^{\prime}=\Omega/(\hbar \omega)$, $\delta^{\prime}=
\delta/(\hbar \omega)$, $\omega_{r}^{\prime}=\omega_{r}/\omega$, $\hat{L}_z=-i \big( x^{\prime} \partial_{y^{\prime}}
-y^{\prime} \partial_{x^{\prime}} \big)$, 
$g_1=N G_1/(\hbar \omega a_{0}^2)$, $g_2=N G_2/(\hbar \omega a_{0}^2)$
and $g_{12}=N G_{12}/(\hbar \omega a_{0}^2)$. 
 
  In simulations for the rotating system we consider $^{87}$Rb atoms and we
use the experimentally-relevant parameters: 
$\lambda=804.1\ \rm nm$, $\omega=2 \pi \times 50\ \rm Hz$ and $\gamma=1$. These parameters give $a_0=\sqrt{\hbar/{m \omega}}=1.52\ \rm{\mu m}$, $k_L^{\prime}=8.42$. From now on we express length in units of $a_0$
(coordinates $(x,y)$ in figures are also given in the units of $a_0$).  
 We performed simulations specifically for the rotation frequency $\omega_r=0.7\ \omega$ and for three different
coupling strengths: no coupling ($\Omega=0$), weak coupling ($\Omega=2\ E_L$), and strong
coupling ($\Omega=10\ E_L$) ($E_L=35.4\ \hbar \omega$). 
  In simulations we choose $g_1=1000$, $g_2=995$, $g_{12}=995$. The ratio between
$g_1$, $g_2$ and $g_{12}$ corresponds to interaction coefficients in $^{87}$Rb (the interaction
coefficients for $^{87}$Rb in states $\lbrace \lvert F=1, m=0 \rangle, \lvert F=1, m=-1 \rangle \rbrace$ is given in Ref.~\cite{Lin2011}). In absence of rotation and for $\Omega=0$, $\delta=0$ 
our choice of $g_1$, $g_2$ and $g_{12}$ produces cloud radius of $8.4\ \rm{\mu m}$.
  We also set $\delta-\hbar \omega_r=0$.
  
  Without rotation and spin-orbit coupling, $\lvert \uparrow \rangle$ and $\lvert \downarrow \rangle$ components are
miscible for our choice of interaction parameters. In the case of rotation and no spin-orbit coupling 
there are several different phases depending on $\omega_r$ and ratio of interaction coefficients \cite{Kasamatsu2003}: triangular lattice, square
lattice, stripe or double-core vortex lattice and vortex sheet. Since our Hamiltonian is almost equivalent to the Hamiltonian in Ref. \cite{Kasamatsu2003} for $\Omega=0$ and $\delta-\hbar \omega_r=0$ (there is a very small difference in interaction coefficients; the equivalence of non-interaction part of two systems is clear from (\ref{app:scheme1_RF_2})) we reproduced results of Ref. \cite{Kasamatsu2003}.

\begin{figure}
\centerline{
\mbox{\includegraphics[width=0.5\textwidth]{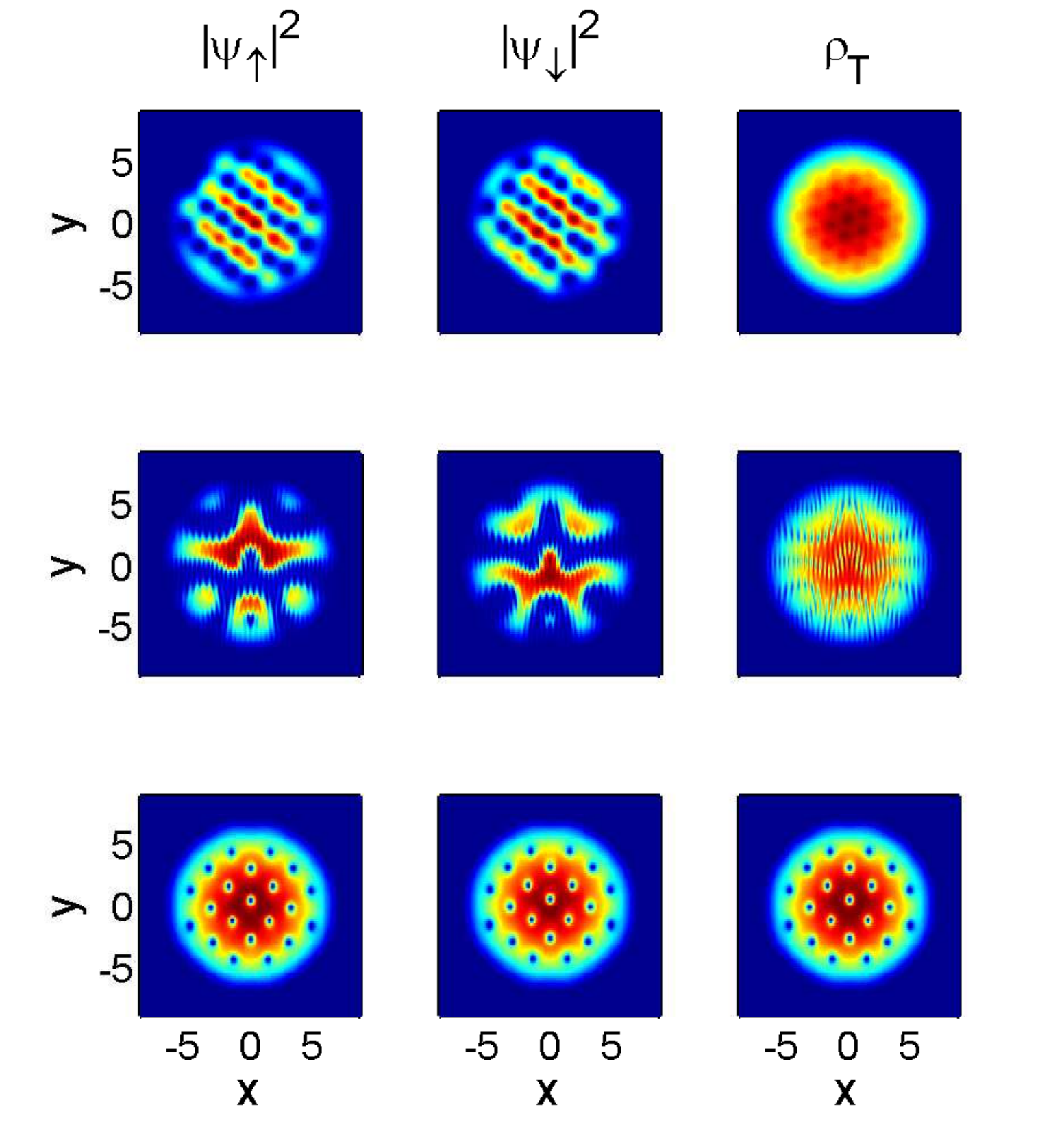}}
\put(-250,253){(a)}
\put(-250,170){(b)}
\put(-250,87){(c)}
}
\caption{
(color online) 
The density profiles for the rotating spin-orbit-coupled BEC are shown. The first, second and third columns
show density of $\lvert \uparrow \rangle$ component ($|\psi_{\uparrow}|^2$), density of 
$\lvert \downarrow \rangle$ component ($|\psi_{\downarrow}|^2$)
and the total density ($\rho_T=|\psi_{\uparrow}|^2+|\psi_{\downarrow}|^2$), respectively. 
Figure (a) shows results for $\Omega=0$ which are characterized by density stripes
and lines of vortices in both components. The results for $\Omega=2\ E_L$ (b) are 
qualitatively similar to the $\Omega=0$ case. Figure (c) shows results for $\Omega=10\ E_L$;
a vortex lattice is formed in both components and densities of the two components are almost identical.
}
\label{rotation}
\end{figure}

The results for $\Omega=0$ are shown in Fig.~\ref{rotation}(a), which display 
the densities of the $\lvert \uparrow \rangle$ and $\lvert \downarrow \rangle$ components
forming spatially-separated density stripes with lines of vortices along the minima of the density. 
As expected, our results reproduce stripe vortex lattice phase described in Ref.~\onlinecite{Kasamatsu2003}.
Note that for $\Omega=0$, the Hamiltonian (\ref{eq:Nature_final})
conserves number of the $\lvert \uparrow \rangle$ and $\lvert \downarrow \rangle$ particles separately.
We have chosen $N_{\uparrow}=N_{\downarrow}$ ($N_i=\int d^2r |\psi_i|^2$). 

A weak spin-orbit coupling ($\Omega=2\ E_L$) (Fig.~\ref{rotation}(b)) does not appear to lead to any significant qualitative changes in the observed behavior: the densities of the $\lvert \uparrow \rangle$ and $\lvert \downarrow \rangle$ 
components are still spatially separated and there are lines of vortices along the density minima of each component.

A significant change comes in the strong-coupling regime: see the $\Omega=10\ E_L$ data shown in Fig. \ref{rotation}(c).
The vortices arrange themselves in a lattice in $\lvert \uparrow \rangle$ and $\lvert \downarrow \rangle$ 
components and densities of both components are almost identical. This behavior can be understood from the following part of the Hamiltonian (\ref{eq:Nature_final}):
\begin{equation}
\hat{H}^{\prime}= \frac{\hbar^2 \hat{k}_{x}^2}{2m} \check{1} 
+ \frac{\hbar^2 k_L \hat{k}_x}{m} \check{\sigma}_z + \frac{\Omega}{2} \check{\sigma}_x.
\label{H_prime_rotation}
\end{equation}
The spectrum of (\ref{H_prime_rotation}) for different $\Omega$'s is shown in Fig. \ref{spectrum}(a).
For large $\Omega$, it consists of two bands with an energy separation
much larger than all other characteristic energies of the system. Therefore, our system is 
``confined'' to the lower band with a single minimum, which effectively makes it a single-component system. This 
explains almost identical densities of the two components in Fig.~\ref{rotation}(c). 

\section{Creating vortices by spatially-dependent detuning}
\label{Sec:detuning}

\subsection{The model}

Vortices in spin-orbit systems like \cite{Lin2011} can be created  without any rotation, but
by imposing an additonal synthetic magnetic field. In \cite{Spielman2009}, it has been shown
that a spatially-dependent detuning, $\delta$, in the M-scheme results in a synthetic 
magnetic field, which creates vortices in the strong Raman coupling ($\Omega$) regime.
Our goal is to investigate the same system for a wide range of $\Omega$ (from weak to strong 
Raman coupling) and to see what kind of vortex structures it yields.
  
 The setup is described by the following effective Hamiltonian (see \cite{Lin2011,Spielman2009}):
\begin{equation}
\hat{H}=\left( \frac{\hbar^2 \hat{{\mathbf k}}^2}{2m} + V 
\right) \check{1} 
+ \frac{\hbar^2 k_L \hat{k}_x}{m} \check{\sigma}_z + \frac{\Omega}{2} \check{\sigma}_x
+ \frac{\delta(y)}{2} \check{\sigma}_z.
\label{H_detuning2}
\end{equation}

 We again assume strong confinement in the $\hat{z}$ direction and describe interactions by
equation (\ref{H_int}).
  We are looking for the ground state in the same way as in the rotating case and following
the same steps we get  the ``dimensionless GP equations'':
\begin{multline*}
\mu \psi_{\uparrow}=
\bigg[-\frac{1}{2} \nabla^{\prime 2} - i k_L^{\prime} \frac{\partial}{\partial x^{\prime}}
+ \frac{1}{2} (x^{\prime 2} + \gamma^2 y^{\prime 2}) \\
+ \frac{\delta^{\prime}(y^{\prime})}{2}
+ g_1 |\psi_{\uparrow}|^2 + g_{12} |\psi_{\downarrow}|^2
\bigg] \psi_{\uparrow} + \frac{\Omega^{\prime}}{2} \psi_{\downarrow}
\end{multline*}
\begin{multline}
\mu \psi_{\downarrow}=
\bigg[-\frac{\hbar^2}{2m} \nabla^{\prime 2} + i k_L^{\prime} \frac{\partial}{\partial x^{\prime}}
+ \frac{1}{2} (x^{\prime 2} + \gamma^2 y^{\prime 2}) \\
- \frac{\delta^{\prime}(y^{\prime})}{2} 
+ g_2 |\psi_{\downarrow}|^2 + g_{12} |\psi_{\uparrow}|^2
\bigg] \psi_{\downarrow} + \frac{\Omega^{\prime}}{2} \psi_{\uparrow}.
\label{dimensionless_GP2}
\end{multline}
Parameters $\Omega^{\prime}$, $\delta^{\prime}$, $k_L^{\prime}$, $g_1$, $g_2$, $g_{12}$ are 
defined in the same way as in (\ref{dimensionless_GP}). 

\subsection{Qualitative discussion}
\begin{figure}[ht]
\centering

\mbox {\subfigure[]{
   \includegraphics[scale =0.29] {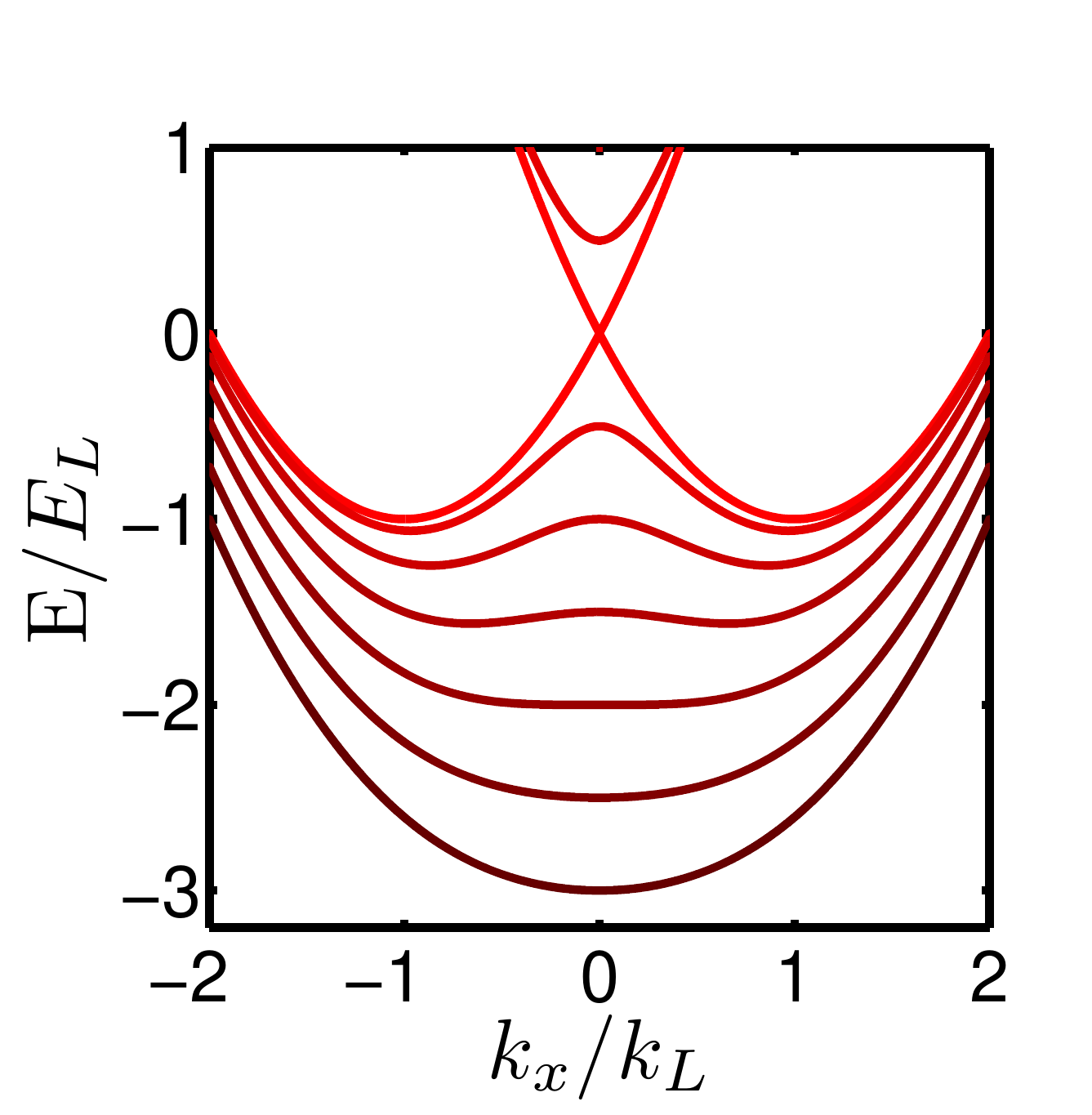}
   \label{spectrum:a}
 }

 \subfigure[]{
   \includegraphics[scale =0.29] {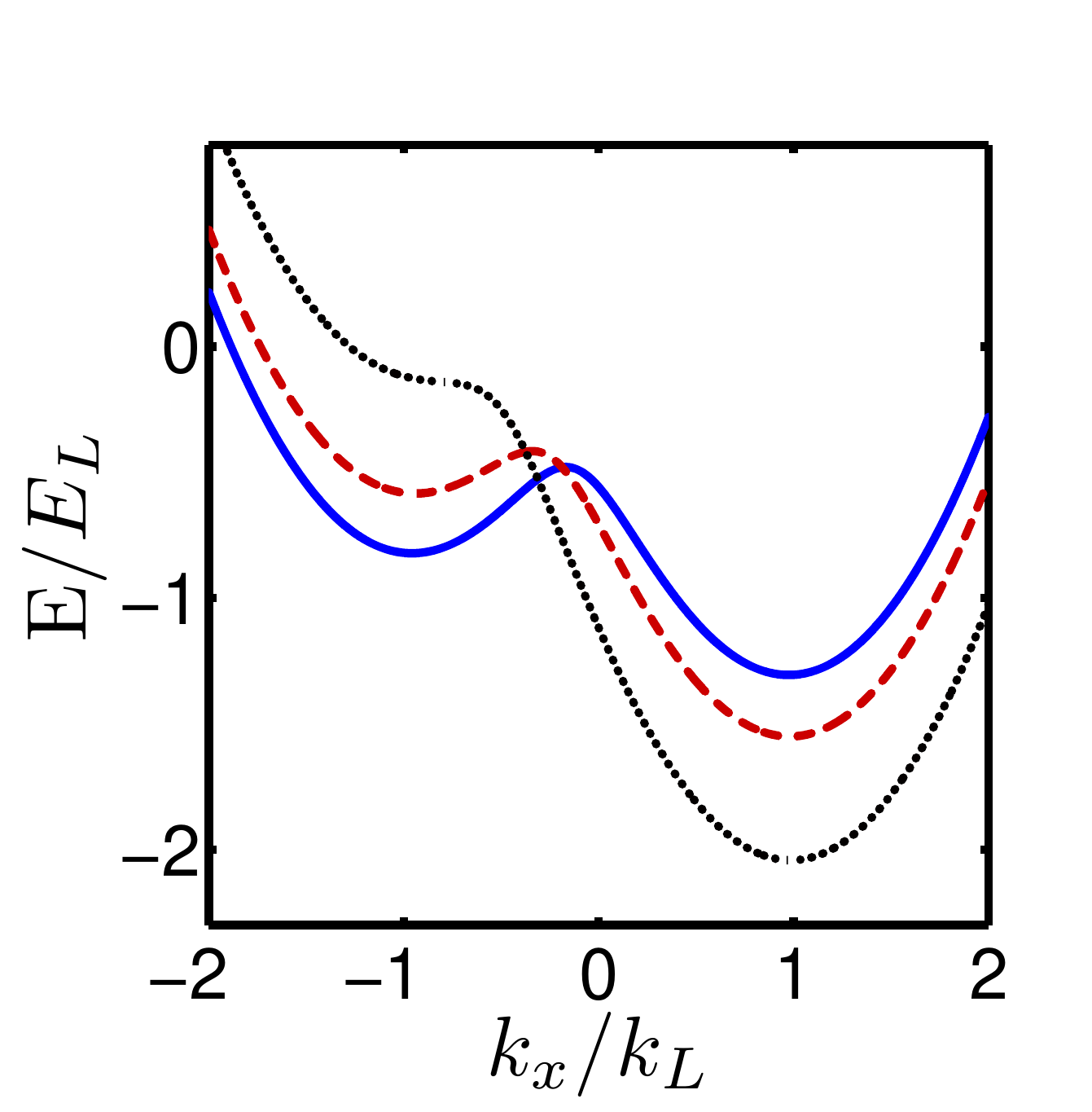}
   \label{spectrum:b}
 }}

 \subfigure[]{
   \includegraphics[scale =0.29] {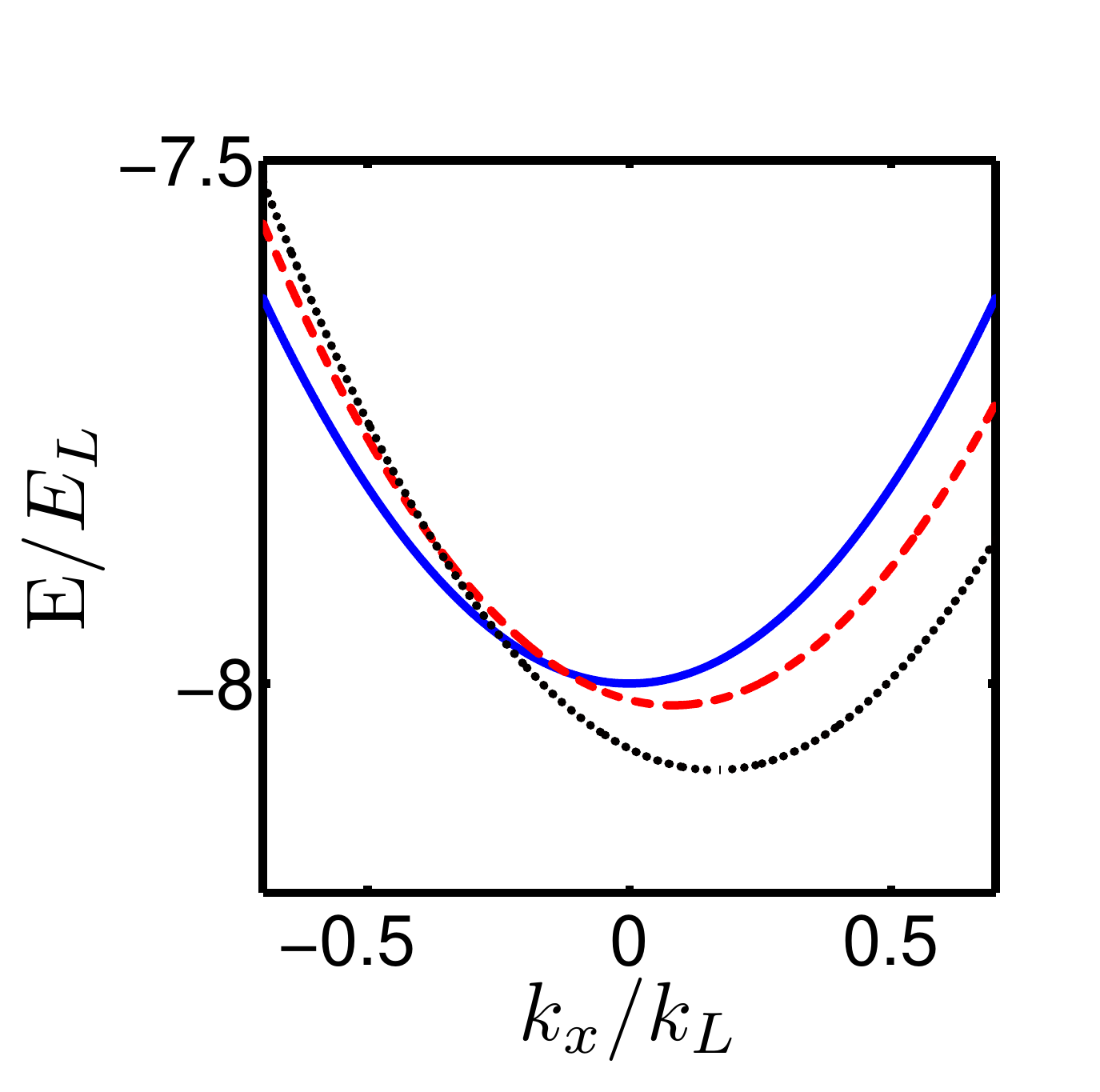}
   \label{spectrum:c}
 }
 
\caption{(color online) The energy spectrum of $H^{\prime}$. In (a) spectra
for different $\Omega$ (from $\Omega=0$ to $\Omega=6\ E_L$) and $\delta=0$ are shown
(spectrum for $\Omega=0$ is at the top while spectrum for $\Omega=6\ E_L$ is at the 
bottom).  
The effect of $\delta$ in small $\Omega$ regime is shown in (b) ($\Omega=1\ E_L$, 
$\delta=0.5\ E_L$ (solid blue line), $\delta=1\ E_L$ (dashed red line) and 
$\delta=2\ E_L$ (dotted black line)). The effect of $\delta$
in large $\Omega$ regime is shown in (c) ($\Omega=16\ E_L$, $\delta=0\ E_L$ 
(solid blue line), $\delta=1\ E_L$ (dashed red line) and $\delta=2\ E_L$ (dotted black line)). 
$\delta$ changes position and energy of the minimum.
}
\label{spectrum}
\end{figure}

  To get a better understanding of the model, we investigate Hamiltonian (\ref{H_detuning2}) 
in more detail. It is instructive to first focus on the following part of (\ref{H_detuning2}):
\begin{equation}
\hat{H}^{\prime}= \frac{\hbar^2 \hat{k}_{x}^2}{2m} \check{1} 
+ \frac{\hbar^2 k_L \hat{k}_x}{m} \check{\sigma}_z + \frac{\Omega}{2} \check{\sigma}_x
+ \frac{\delta}{2} \check{\sigma}_z.
\label{H_prime}
\end{equation}
We first assume that $\delta$ is constant in space. In that case Hamiltonian
(\ref{H_prime}) can be easily diagonalized in the momentum basis:
$U^{\dagger}(k_x)H^{\prime}(k_x)U(k_x)=
\bigg(\begin{smallmatrix}
E_{+}(k_x)&0 \\ 0&E_{-}(k_x)
\end{smallmatrix} \bigg)$. 
  The resulting spectrum consists of an upper(+) and lower(-) band, as shown in 
Fig. \ref{spectrum}. The gap separating the bands is large compared to other characteristic 
energies of the system and it is safe to assume that the condensate occupies only the 
states in the lower band. In Fig. \ref{spectrum}(a), spectra for different coupling strengths 
$\Omega$ and $\delta=0$ are shown. For $\Omega<4 E_L$, the spectrum has two 
minima and BEC will involve states near both left and right minima.
At $\Omega=4 E_L$, there is a transition from a spectrum with
two minima to a spectrum with one minimum, which changes the structure of 
the condensate wave-function. I.e., for $\Omega>4 E_L$, the BEC is expected to occupy 
only states with momentum around $k_x=0$. 
 
 The effect of detuning $\delta$ in the low-$\Omega$ regime is shown in 
Fig. \ref{spectrum}(b). We see that $\delta$ shifts the energies and positions of the left 
and right minima. In the case of constant $\delta$, the  BEC would occupy only the states around
the global minimum (for example, the right minimum in Fig. \ref{spectrum}(b)). Those cases
have been tested experimentally in \cite{Lin2011}. 

Now, consider a spatially-dependent $\delta(y)$. We will consider it to be a linear
function of $y$: $\delta(y)=\delta_0 + \beta y$, which is the simplest and the most experimentally 
relevant regime.  The interesting physics is evident from the following arguments: for 
constant detuning, the spectrum around a minimum can be simply described by (we use dimensionless variables, see 
(\ref{dimensionless_GP})) 
$(k_{x}-k_{\rm min})^2/(2 m_{\rm eff})  + 
E_{\rm min}$, where $m_{\rm eff}$, $k_{\rm min}$, $E_{\rm min}$ are the effective mass, position of the minimum, and the energy at the minimum, respectively. Note that all these quantities depend on  $\delta$.
  If $\delta$ is $y$-dependent, the values of $m_{\rm eff}$, $k_{\rm min}$, $E_{\rm min}$ will
also become spatially-dependent. Hence, the spectrum around the minimum can now be written 
as: $\big[ k_{x}-k_{\rm min}(y) \big]^2/(2 m_{\rm eff}) + E_{\rm min}(y)$ which describes 
particles moving in an effective gauge field $({\mathbf A},\Phi)=\big( k_{\rm min}(y),0,0,E_{\rm min}(y) \big)$ 
with a spatially-varying effective mass $m_{\rm eff}(y)$ \cite{Spielman2009}. The spatially-dependent vector potential 
${\mathbf A}$ induces an effective magnetic field (${\mathbf B}_{\rm eff}=\bm {\nabla} \times {\mathbf A})$, which may lead to creation of vortices if strong enough.   This approximation provides a good
description of the system only if the particles at some point $y$ have the momentum $k_x$ near
the minimum. Our numerical simulations presented below indicate that this approximation in fact gives a very good qualitative description in a wide parameter range. 
  
 We calculate parameters $m_{\rm eff}(y)$, $k_{\rm min}(y)$,  and $E_{\rm min}(y)$ by diagonalizing (\ref{H_prime}) for different $y$'s since $\delta=\delta(y)$.
  The procedure of deriving effective equations for lower band for Hamiltonian 
(\ref{H_detuning2}) in high $\Omega$ (single minimum) regime is described in 
Ref.~[\onlinecite{Spielman2009}]. Let us note however, that the method we use below to find 
the ground state is exact (in particular, we do not limit our system to lower band 
and we do not simplify interaction terms).

\subsection{Results}
\label{Sec:Results}

  In simulations for a system with a spatially-dependent detuning $\delta$ we
use the same experimental parameters as in the simulations of a rotating system,
which gives $a_0=\sqrt{\hbar/(m \omega)}=1.52\ \rm{\mu m}$ and 
$k_L^{\prime}=8.42$. We choose interaction parameters to be $g_1=1600$, $g_2=1593$, 
$g_{12}=1593$ and constant part of detuning $\delta_0=0$.
  
\begin{figure}[ht]
\centering

\subfigure[]{
   \includegraphics[scale =0.4] {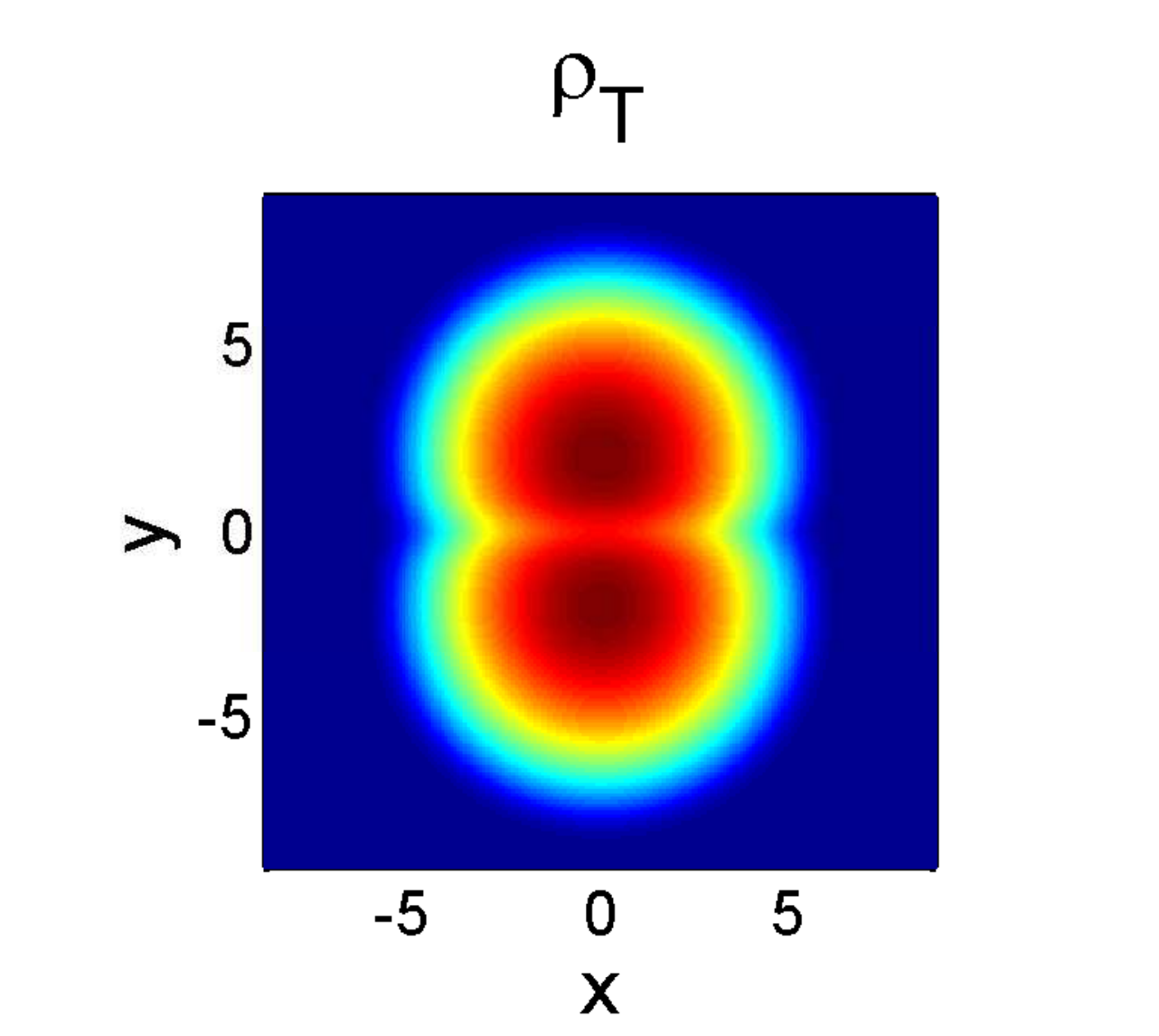}
   \label{Omega0:a}
 }

  \subfigure[]{
   \includegraphics[scale =0.39] {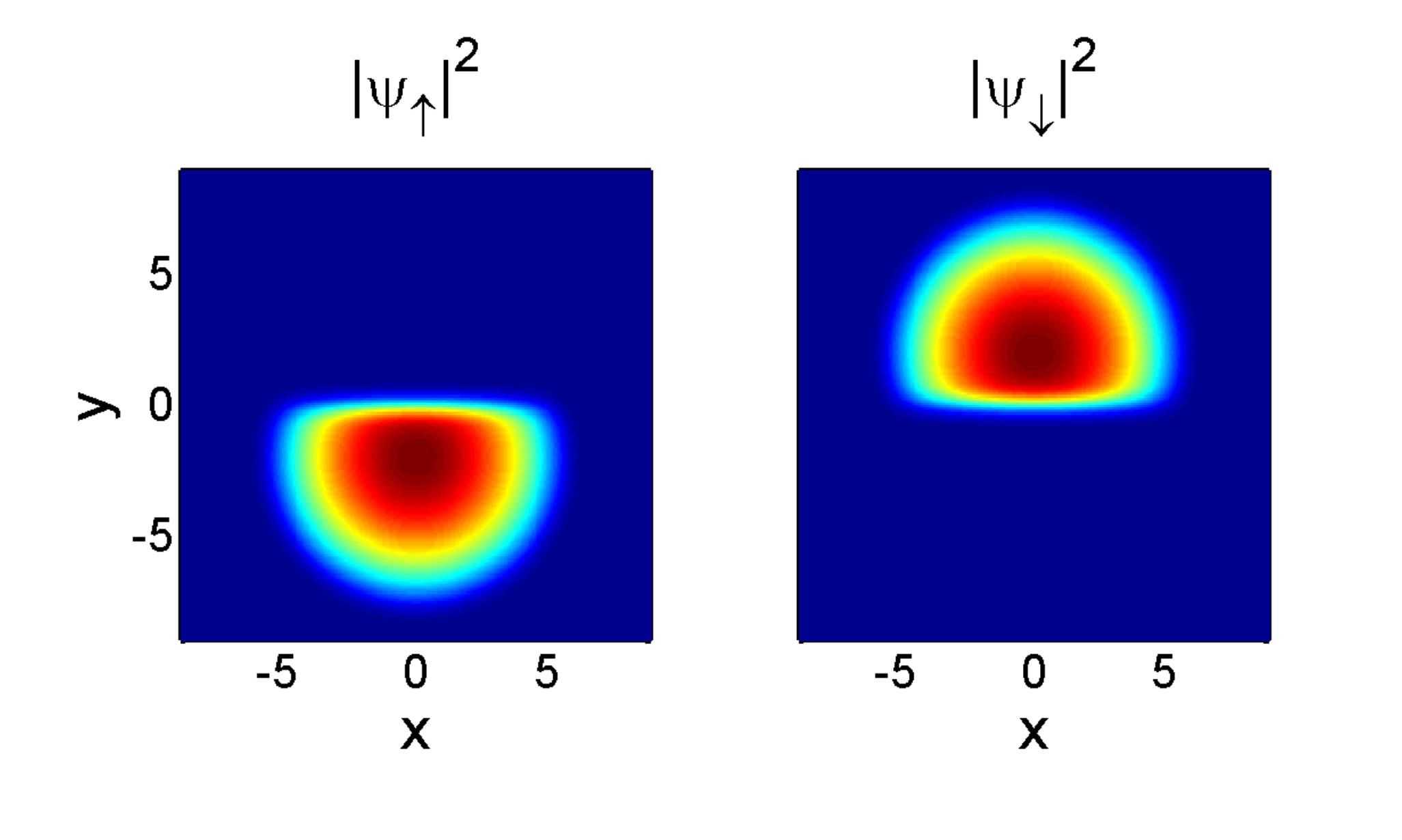}
   \label{Omega0:b}
 }
 \subfigure[]{
   \includegraphics[scale =0.39] {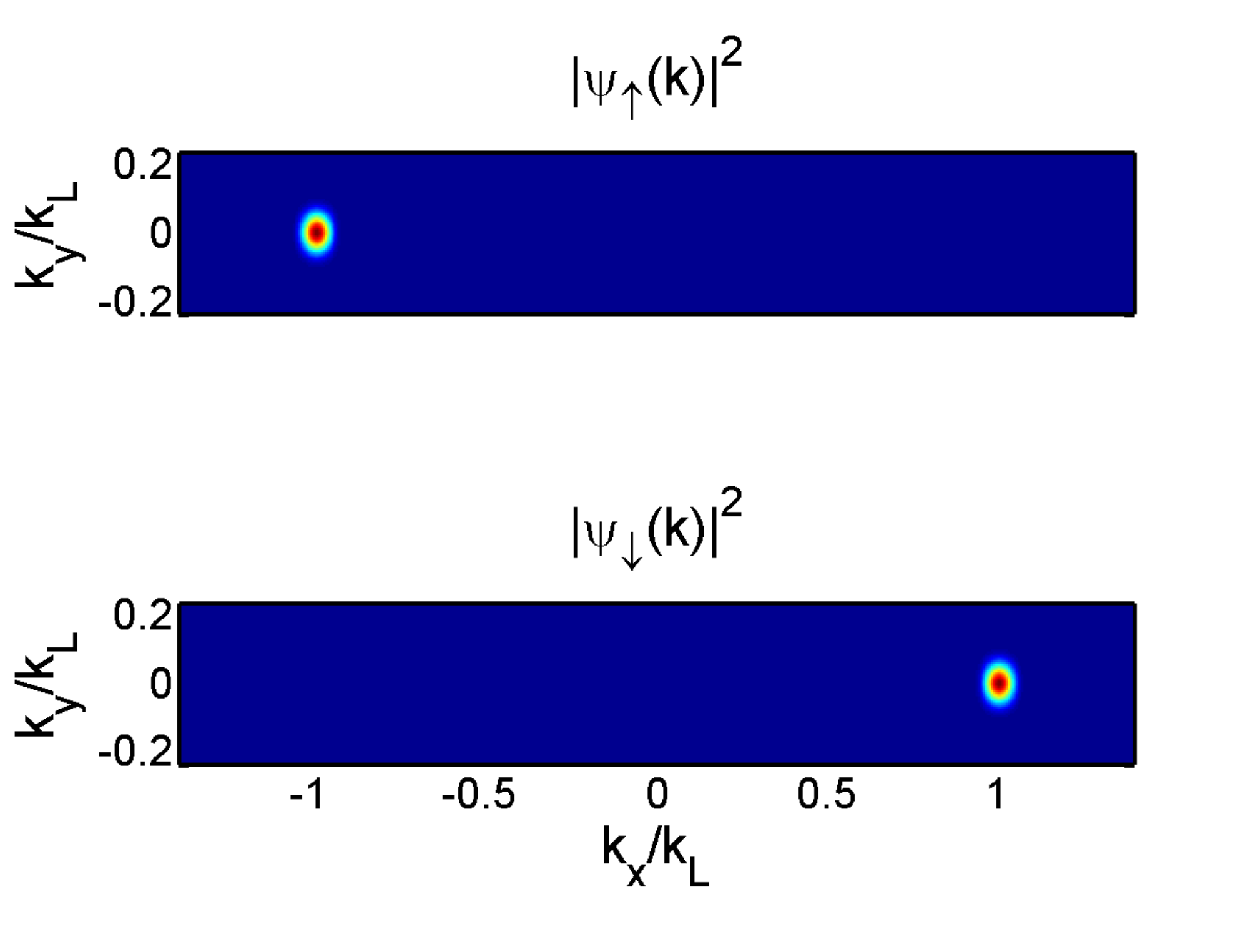}
   \label{Omega0:c}
 }
 
\caption{(color online) 
The figure shows results for $\Omega=0$, $\beta=4\ \hbar \omega/a_0$ and $\gamma=1$. 
In (a) the total density is shown. The shape of the density is determined
by spatially-dependent detuning, which shifts the densities of $\lvert \uparrow \rangle$
and $\lvert \downarrow \rangle$ particles (b). Momentum distribution of 
$\lvert \uparrow \rangle$ and $\lvert \downarrow \rangle$ components is shown in
(c).
}
\label{Omega0}
\end{figure}

  The results for $\Omega=0$, $\beta=4\ \hbar \omega/a_0$ and $\gamma=1$ are shown in 
Fig.~\ref{Omega0} and are straightforward to understand. In this case, we may write the 
Hamiltonian (\ref{H_detuning2})  as 
$$
\hat{H}=
\big( \begin{smallmatrix}
H_{\uparrow}&0 \\ 0&H_{\downarrow}
\end{smallmatrix} \big),
$$
where 
$H_{\uparrow}=\frac{\hbar^2}{2m} \left( \hat{{\mathbf k}}^2 + 2 k_L \hat{k}_x \right)
+ V_{\uparrow}({\mathbf r})$, $H_{\downarrow}=\frac{\hbar^2}{2m} 
\left( \hat{{\mathbf k}}^2 - 2 k_L \hat{k}_x \right) + V_{\downarrow}({\mathbf r})$
and 
$V_{\uparrow}({\mathbf r})=V({\mathbf r})+
\delta(y)/2$, $V_{\downarrow}({\mathbf r})=V({\mathbf r})
-\delta(y)/2.$
  We see that motion of $\lvert \uparrow \rangle$ and $\lvert \downarrow \rangle$
particles is decoupled in $\hat{H}$ and that they experience different potentials
$V_{\uparrow}({\mathbf r})$, $V_{\downarrow}({\mathbf r})$. Detuning gradient $\beta$ 
shifts the minima of 
$V_{\uparrow}({\mathbf r})$ ($V_{\downarrow}({\mathbf r})$) for $y_0=\beta/( 2 m \omega^2 
\gamma^2)$ in the positive/negative $\hat{y}$-direction and 
therefore, the centers of the $\lvert \uparrow \rangle$ and $\lvert \downarrow 
\rangle$ densities are shifted from the origin by $\pm y_0$ (the origin is located in the 
minimum of $V({\mathbf r})$), see Fig. \ref{Omega0}(b). Also, it is clear from 
$\hat{H}$ and Fig. \ref{spectrum}(a) that the momentum distribution
of $\lvert \uparrow \rangle$ ($\lvert \downarrow \rangle$) particles will be centered around 
${\mathbf k}=(-k_L, 0)$ (${\mathbf k}=(k_L, 0)$), see Fig. \ref{Omega0}(c). 
  The effect of repulsive interactions between the particles with 
different spins is clearly seen (the overlap between $\lvert \uparrow \rangle$ and 
$\lvert \downarrow \rangle$ densities is quite small). 

\begin{figure}[ht]
\centering

\subfigure[]{
   \includegraphics[scale =0.38] {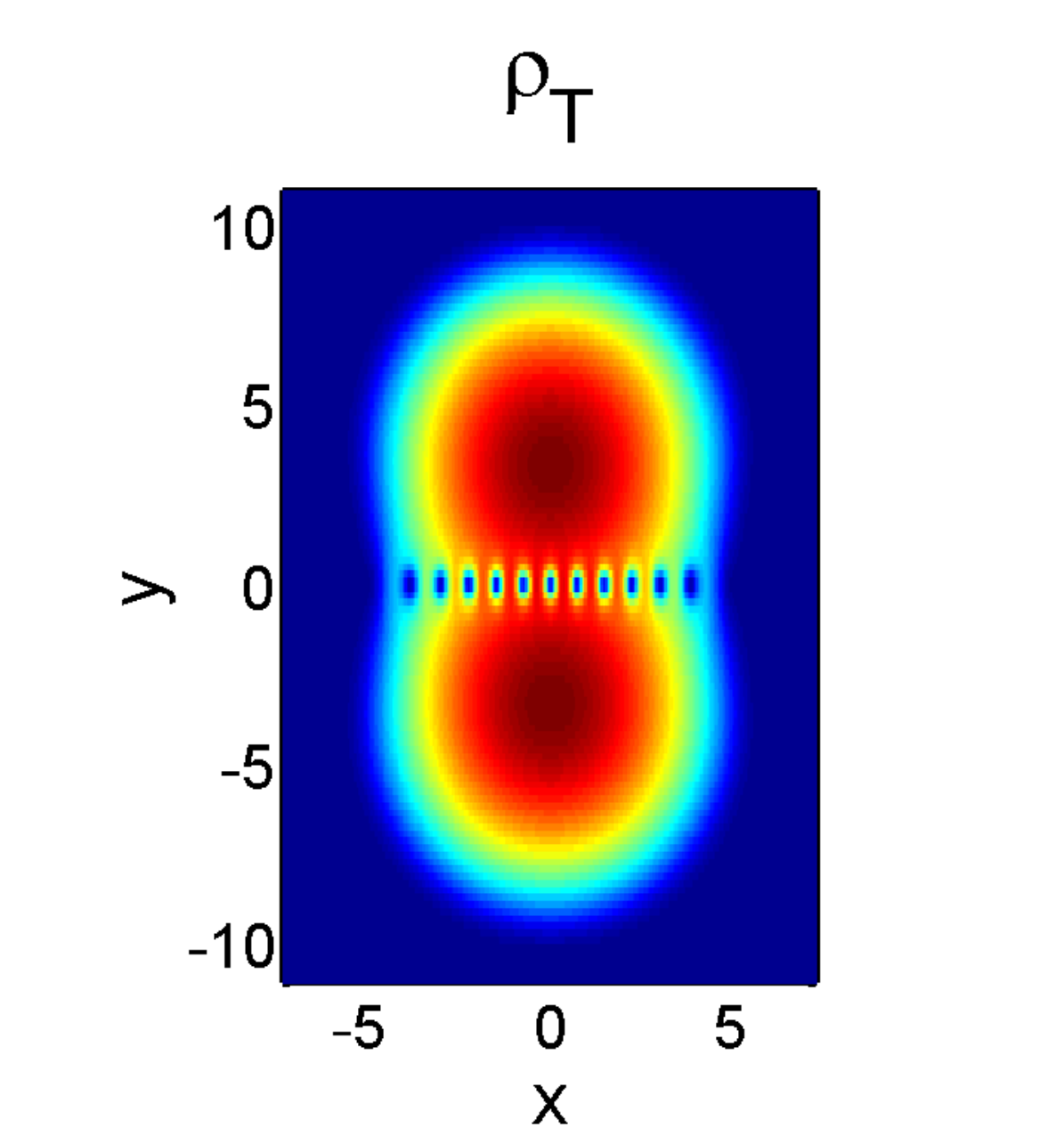}
   \label{Omega5:a}
 }

\subfigure[]{
   \includegraphics[scale =0.35] {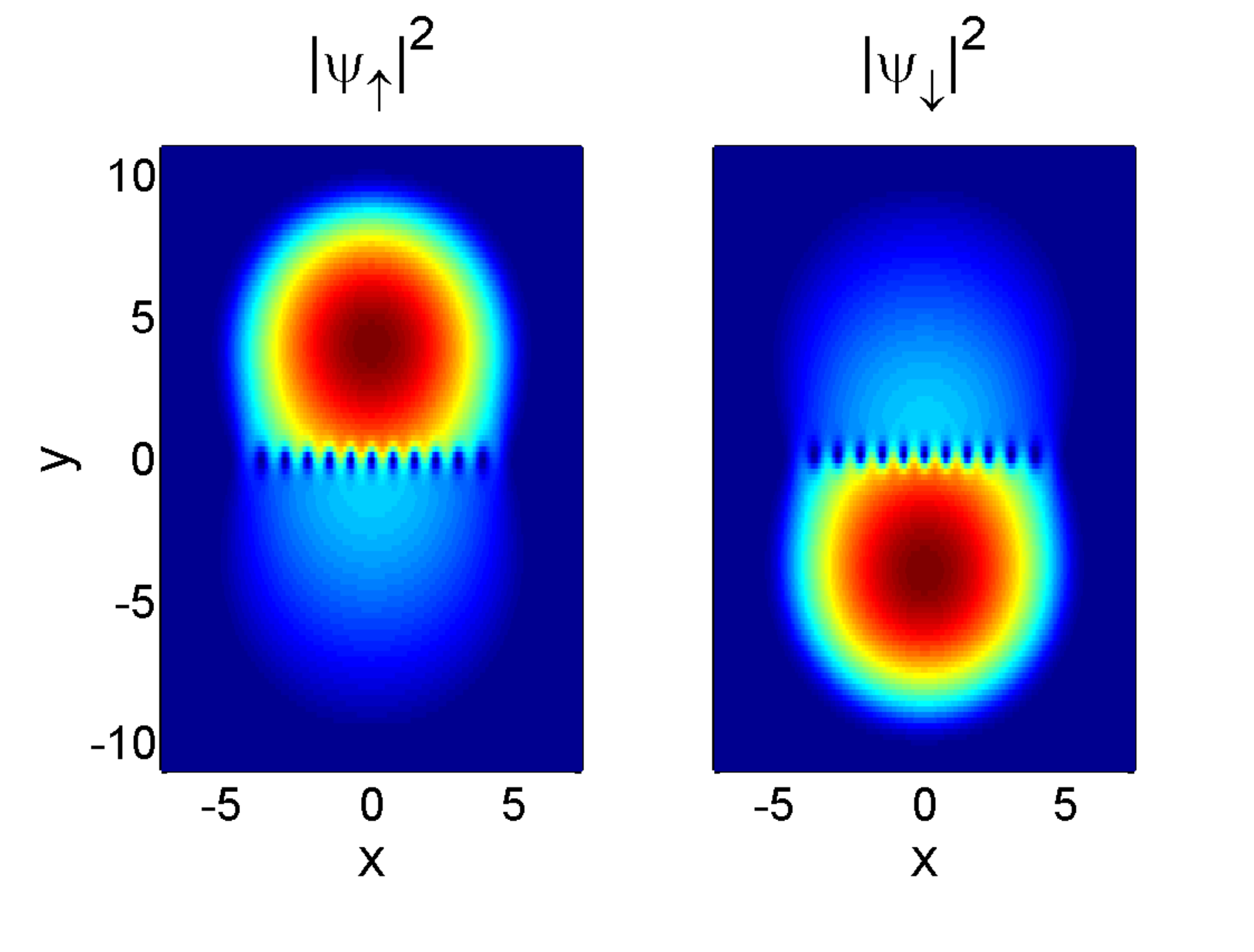}
   \label{Omega5:b}
 } 
 
\subfigure[]{
   \includegraphics[scale =0.35] {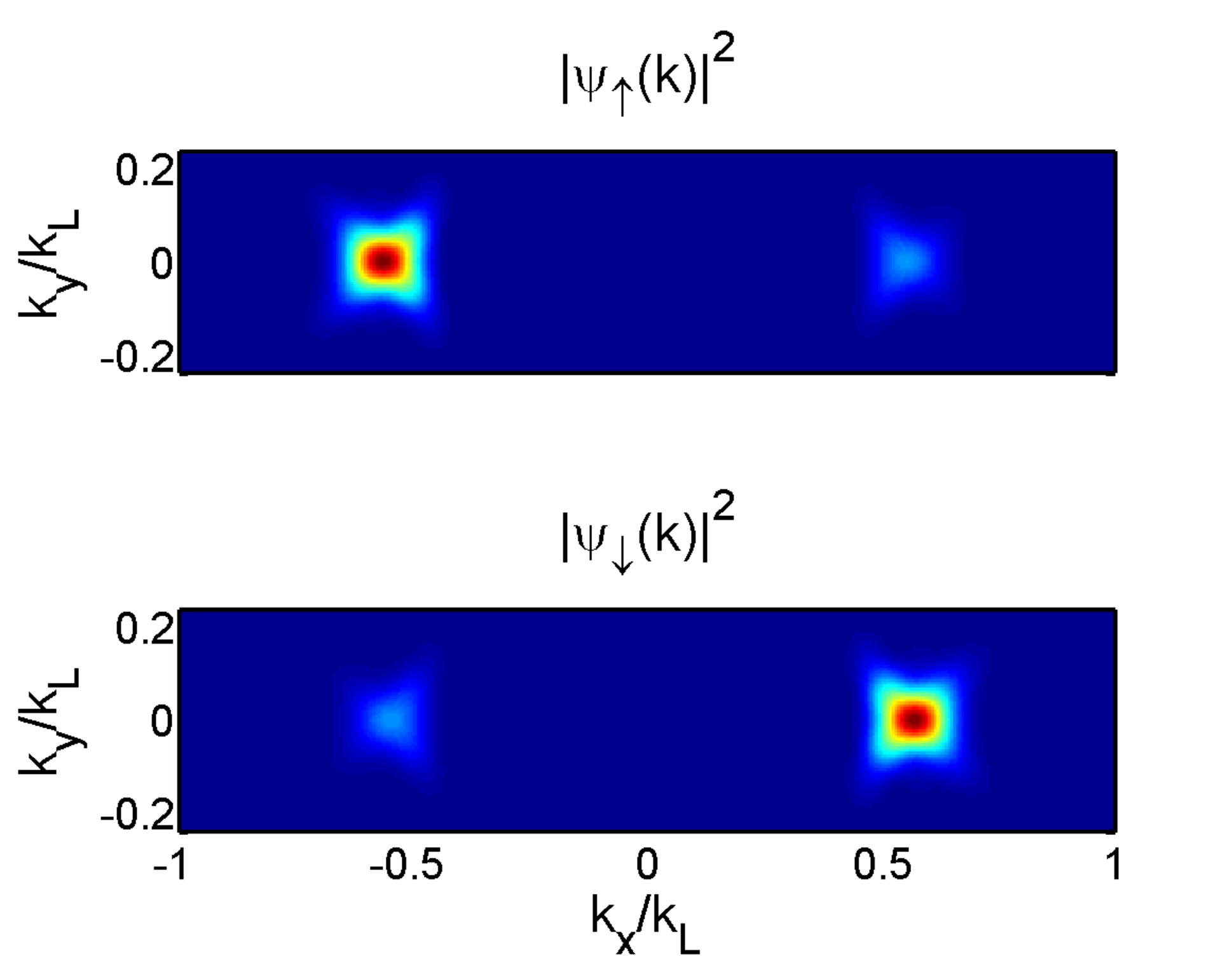}
   \label{Omega5:c}
 } 
 
\caption{(color online)
The figure shows results for $\Omega=5\ E_L$, $\beta=12\ \hbar \omega/a_0$
and $\gamma=1$. 
In (a) the total density is shown. The series of minima at $y=0$ comes
from vortices in $\lvert \uparrow \rangle$ and $\lvert \downarrow \rangle$  
wavefunctions (b). Momentum distribution of 
$\lvert \uparrow \rangle$ and $\lvert \downarrow \rangle$ components is shown in
(c).
}
\label{Omega5}
\end{figure}

  If we introduce a finite $\Omega$, the Hamiltonian becomes: 
$\hat{H}=\bigg( \begin{smallmatrix}
H_{\uparrow}&\Omega/2 \\ \Omega/2&H_{\downarrow}
\end{smallmatrix} \bigg)$, 
The corresponding $\Omega$-term creates coupling between $\lvert \uparrow \rangle$ 
and $\lvert \downarrow \rangle$ particles. If $\delta=\rm {const}=0$ and $\Omega$ is small,
the states around the left (right) minimum in the spectrum in Fig. \ref{spectrum}(a)
still consist mainly of the $\lvert \uparrow \rangle$ ($\lvert \downarrow \rangle$) particles,
but there is also some admixture of the component with the opposite spin, which grows with 
$\Omega$. It means that $\psi_{\uparrow}({\mathbf r})$ ($\psi_{\downarrow}({\mathbf r})$) 
will mainly consist of states with momentum around the left (right) minimum, but also of states 
around the right (left) minimum. We can therefore write:  
\begin{equation}
\begin{pmatrix} 
\psi_{\uparrow}({\mathbf r})\\
\psi_{\downarrow}({\mathbf r}) 
\end{pmatrix}  = 
\begin{pmatrix} 
\psi_{\uparrow L}({\mathbf r})\\
\psi_{\downarrow L}({\mathbf r}) 
\end{pmatrix}  +
\begin{pmatrix} 
\psi_{\uparrow R}({\mathbf r})\\
\psi_{\downarrow R}({\mathbf r}) 
\end{pmatrix},
\end{equation}
where $\psi_{\uparrow L}({\mathbf r})$ and $\psi_{\downarrow L}({\mathbf r})$ consist only of 
states with momenta around left peak, while $\psi_{\uparrow R}({\mathbf r})$ and 
$\psi_{\downarrow R}({\mathbf r})$ consist only of the states with momenta around right peak of 
momentum distribution. We therefore call $\big( \psi_{\uparrow L}({\mathbf r}),
\psi_{\downarrow L}({\mathbf r}) \big)$ and $\big( \psi_{\uparrow R}({\mathbf r}), 
\psi_{\downarrow R}({\mathbf r}) \big)$
left and right wave-function. In the spatially-dependent detuning case it may happen
that momentum distribution is separated in two peaks (i.e., there exist ``left''- and ``right-movers'') even for $\Omega>4\ E_L$ (see for example Fig.~\ref{Omega5}(c)). In that case also the notion of left and right wave-function applies. 
  
\begin{figure}[ht]
\centering

\subfigure[]{
   \includegraphics[scale =0.35] {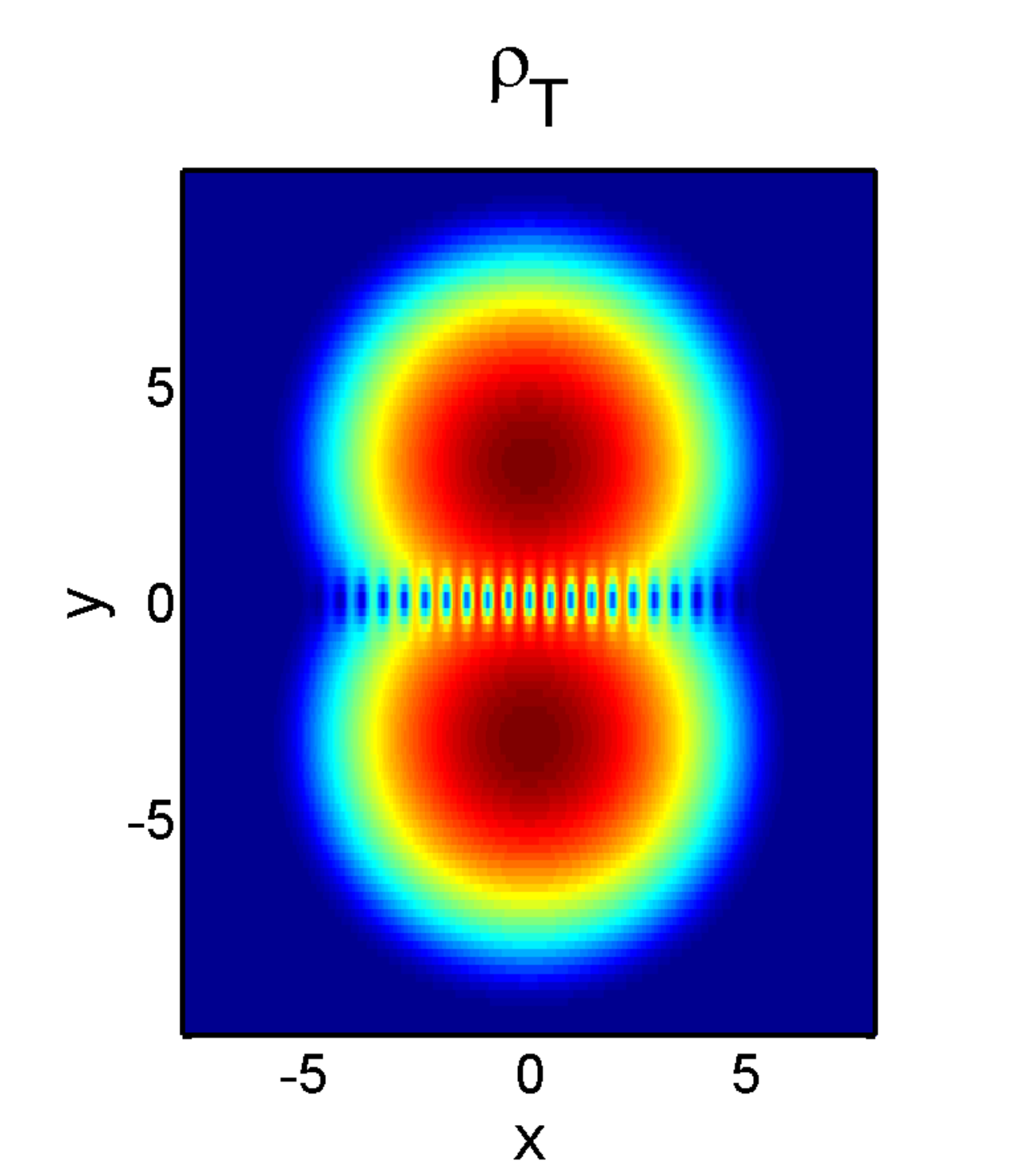}
   \label{Omega3:a}
 }

\subfigure[]{
   \includegraphics[scale =0.35] {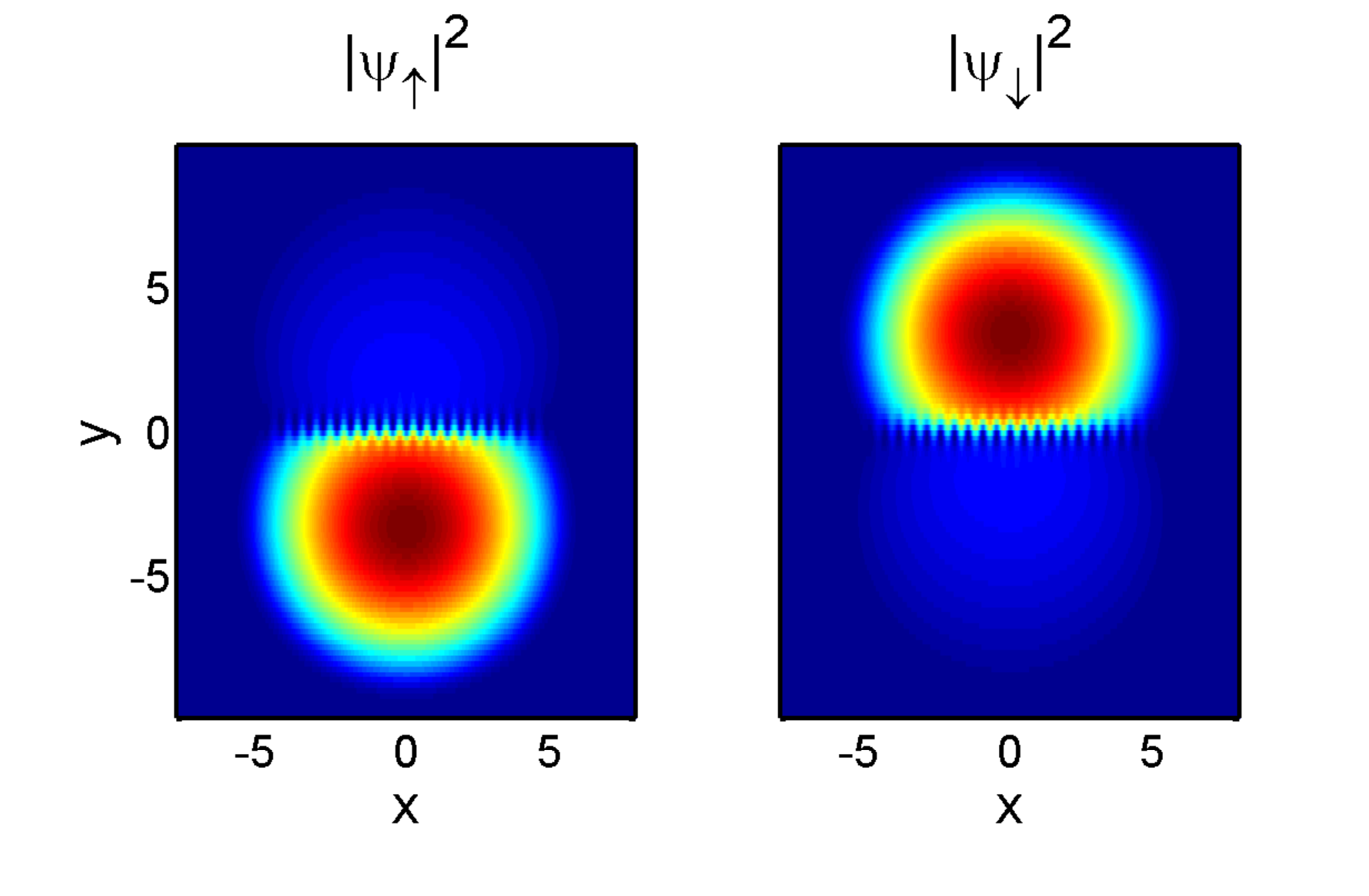}
   \label{Omega3:b}
 } 
 
\subfigure[]{
   \includegraphics[scale =0.35] {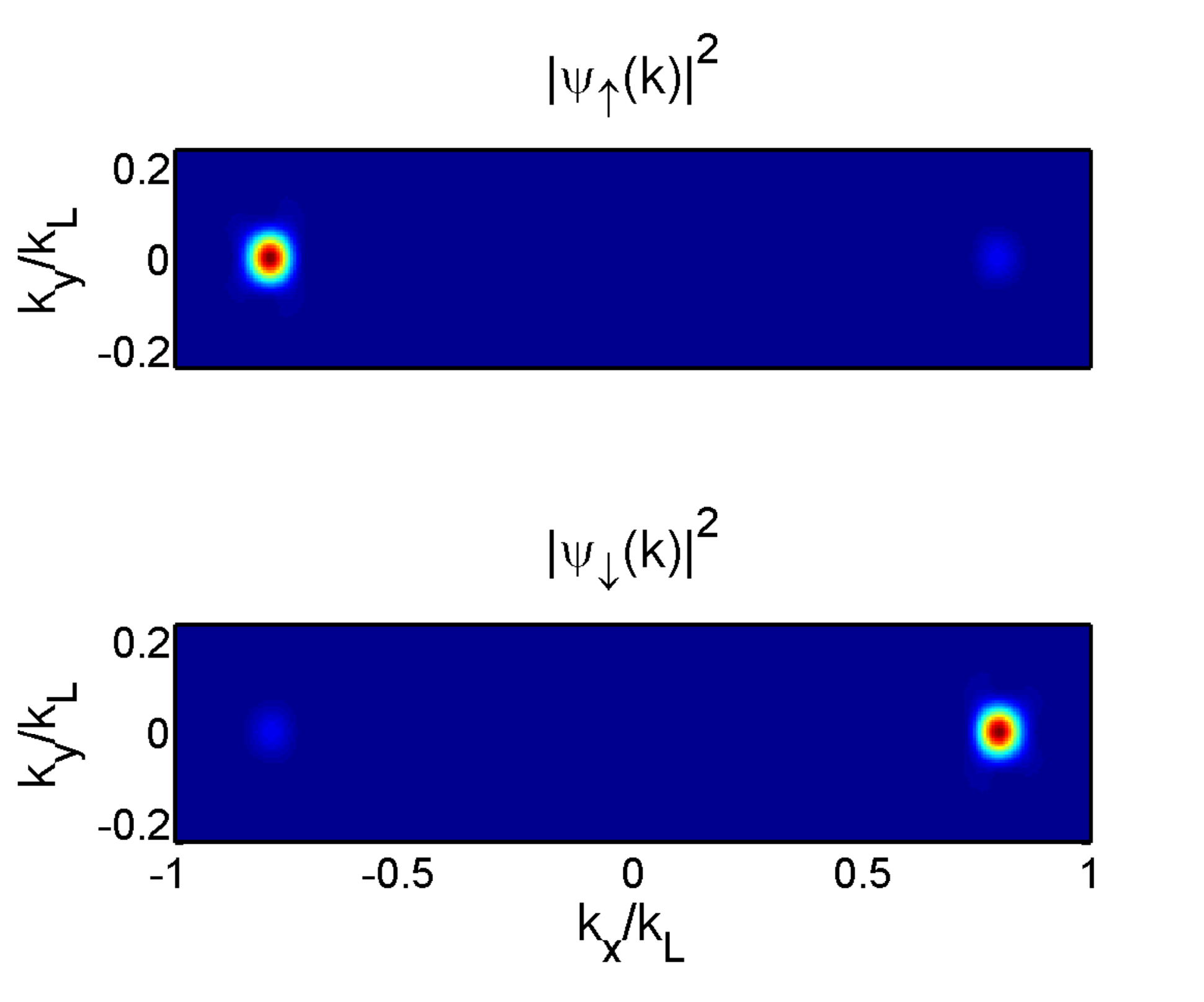}
   \label{Omega3:c}
 } 
 
\caption{(color online)
The figure shows results for $\Omega=3\ E_L$, $\beta=8\ \hbar \omega/a_0$
and $\gamma=1$. 
In (a) the total density is shown. The series of minima at $y=0$ comes
from vortices in $\lvert \uparrow \rangle$ and $\lvert \downarrow \rangle$  
wavefunctions (b). Momentum distribution of 
$\lvert \uparrow \rangle$ and $\lvert \downarrow \rangle$ components is shown in
(c).
}
\label{Omega3}
\end{figure}

\begin{figure}[ht]
\centering

 \subfigure[]{
   \includegraphics[scale =0.4] {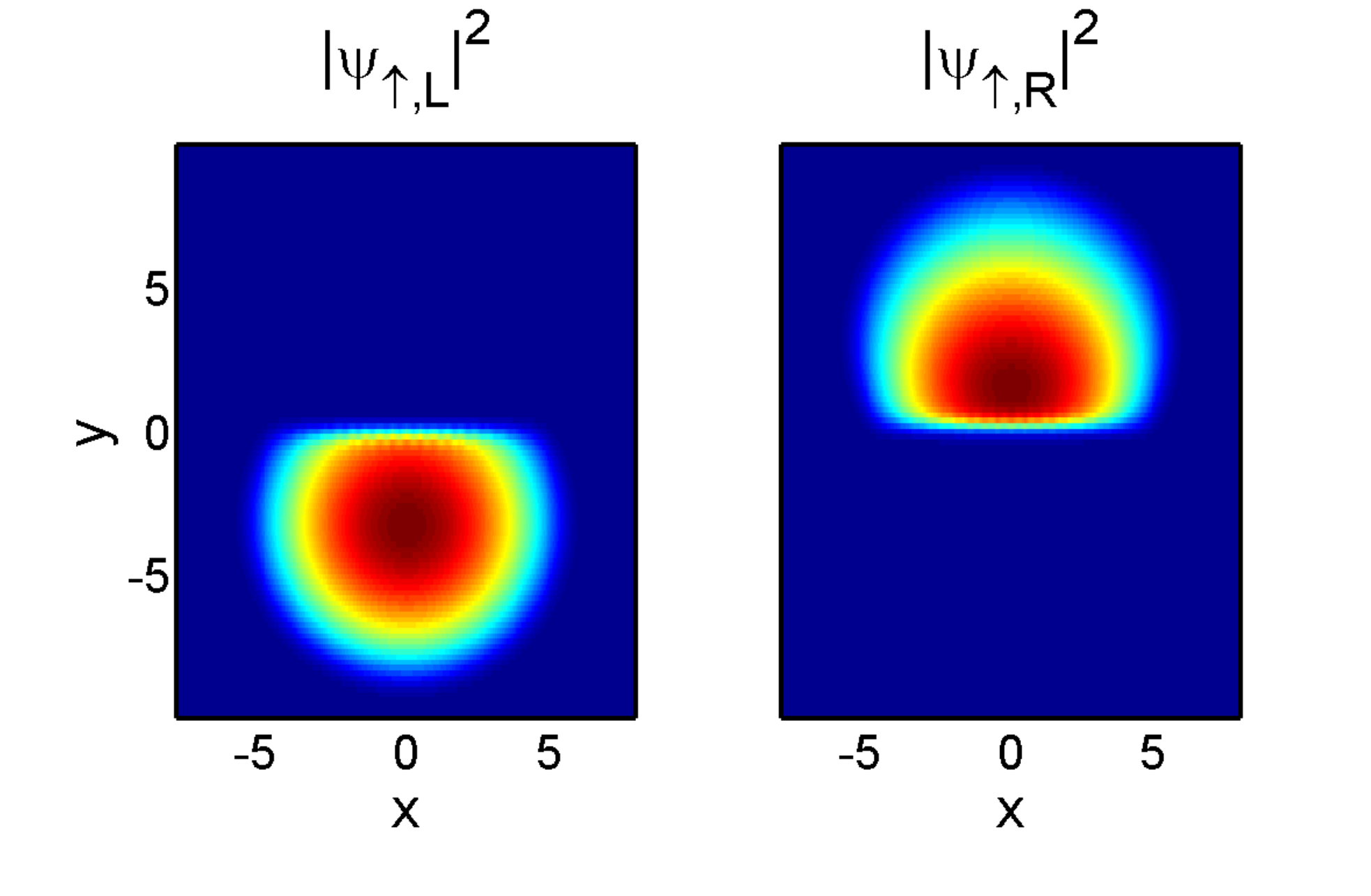}
   \label{Omega3_details:a}
 }
 
  \subfigure[]{
   \includegraphics[scale =0.4] {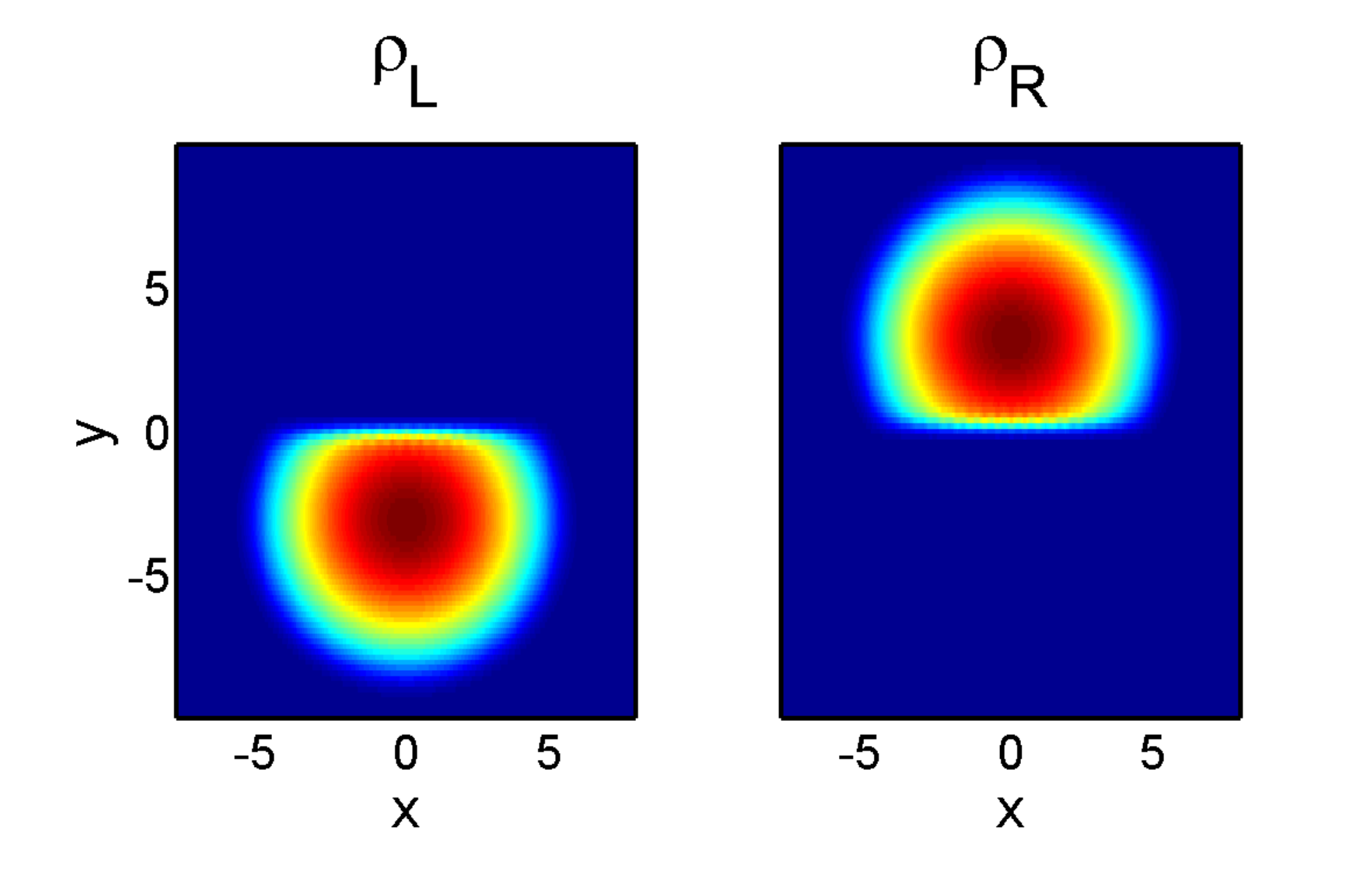}
   \label{Omega3_details:b}
 }

\caption{(color online)
The figure (a) shows $|\psi_{\uparrow L}|^2$ and $|\psi_{\uparrow R}|^2$ 
the relative amplitude of which is given by $\int d^2r |\psi_{\uparrow L}|^2=0.45$ and
$\int d^2r |\psi_{\uparrow R}|^2=0.05$ for the parameters $\Omega=3\ E_L$, 
$\beta=8\ \hbar \omega/a_0$ and $\gamma=1$.
The superposition of $\psi_{\uparrow L}$ and
$\psi_{\uparrow R}$, $\psi_{\uparrow}=\psi_{\uparrow L}+\psi_{\uparrow R}$, produces 
vortices in $\psi_{\uparrow}$. The density of left- and right-moving particles 
($\rho_L=|\psi_{\uparrow L}|^2+|\psi_{\downarrow L}|^2$, $\rho_R=|\psi_{\uparrow R}|^2+
|\psi_{\downarrow R}|^2$) particles is shown in (b).
}
\label{Omega3_details}
\end{figure}

   To investigate the effect of $\Omega$, which couples $\lvert \uparrow \rangle$ and 
$\lvert \downarrow \rangle$ states, we consider the regime with $\Omega=3\ E_L$ and 
$\beta= 8\ \hbar \omega/a_0$ (Fig. \ref{Omega3}). The total density $\rho_T({\mathbf r})$ is 
shown in Fig. \ref{Omega3}(a) and there is a characteristic series of minima along the $x$-direction at
$y=0$, which come from vortices in the $\psi_{\uparrow}$ and $\psi_{\downarrow}$ wave-functions, see
Fig.~\ref{Omega3}(b), which are positioned along $x$ and near $y=0$. We have checked that
the phase winding around zero density points of $|\psi_{\uparrow}|^2$ and 
$|\psi_{\downarrow}|^2$ is $-2 \pi$. Since vortices in
$\lvert \downarrow \rangle$ and $\lvert \uparrow \rangle$ components are slightly displaced from $y=0$, the density at minima in $\rho_T$ are close to, but not exactly equal to zero.
  To explain the existence of the line of vortices in the $\lvert \uparrow \rangle$ and 
$\lvert \downarrow \rangle$ components, we examine the left and right wave-functions. 
Fig. \ref{Omega3_details}(a) displays $|\psi_{\uparrow L}|^2$ and $|\psi_{\uparrow R}|^2$  
(note that the amplitude of $\psi_{\uparrow R}$ is considerably smaller than the amplitude of
$\psi_{\uparrow L}$: $\int d^2r |\psi_{\uparrow R}|^2=0.05$ and $\int d^2r |\psi_{\uparrow L}|^2=0.45$). 
The momentum distribution in Fig.~\ref{Omega3}(c) shows that the wave-packet, $\psi_{\uparrow L}$, has an average momentum of $k_{\rm left}=-0.8\ k_L$  and $\psi_{\uparrow R}$ has an average momentum of $k_{\rm right}=0.8\ k_L$. Since $\psi_{\uparrow}$  is a superposition of  the left-  and right-movers, $\psi_{\uparrow}=\psi_{\uparrow L}+\psi_{\uparrow R}$, the appearance of the line of vortices at overlapping region is expected. The separation of vortices $d$ is then simply given by $(k_{\rm right}-k_{\rm left})d=2\pi$ or $d=(2\pi)/(k_{\rm right}-k_{\rm left})$. 
The analytical expression for $d$ fits perfectly well to our numerical data.

  To explain the density profile and momentum distribution, it is useful to consider an effective gauge-field picture. The effective gauge field, $({\mathbf A},\Phi)
=(k_{\rm min}(y),0,0,E_{\rm min}(y))$, can be calculated by diagonalizing $H^{\prime}$. As discussed earlier, we may approximate the low-energy band physics by the following Hamiltonian (we use again the dimensionless variables, where
the lengths are measured in terms of $a_0$ and the wave-vectors, $k$, in terms of $1/a_0$):
\begin{equation}
H_{\rm EGF}=\frac{1}{2 m_{\rm eff}(y)} \big[ k_x - A(y) \big]^2
+ \frac{1}{2} k_{y}^2 + \Phi(y) + V({\mathbf r}),
\label{H_EGF}
\end{equation}
where $V({\mathbf r})=\frac{1}{2}(x^2+\gamma^2 y^2)$. For $\Omega \geq 4 E_L$ there is a single local minimum in
lower band of the Hamiltonian (\ref{H_prime}) spectrum for any $\delta$. For $\Omega < 4 E_L$ the spectrum has two minima
for $\delta=0$, however when $\delta$ becomes large enough the spectrum has a single local minimum (Fig. \ref{spectrum}(b)). The spectrum around each local minimum can be approximated by the form given in (\ref{H_EGF}),
and therefore there will be $A_L(y)$, $\Phi_L(y)$, $m_{\rm eff,L}(y)$ corresponding to the left minimum and
$A_R(y)$, $\Phi_R(y)$, $m_{\rm eff,R}(y)$ corresponding to the right minimum of the spectrum. Left-movers feel
the ``left gauge field'' $(A_L(y),0,0,\Phi_L(y))$ while right-movers feel the ``right gauge field'' $(A_R(y),0,0,\Phi_R(y))$.
 
\begin{figure}[ht]
\centering

\mbox{\subfigure[]{
   \includegraphics[scale =0.26] {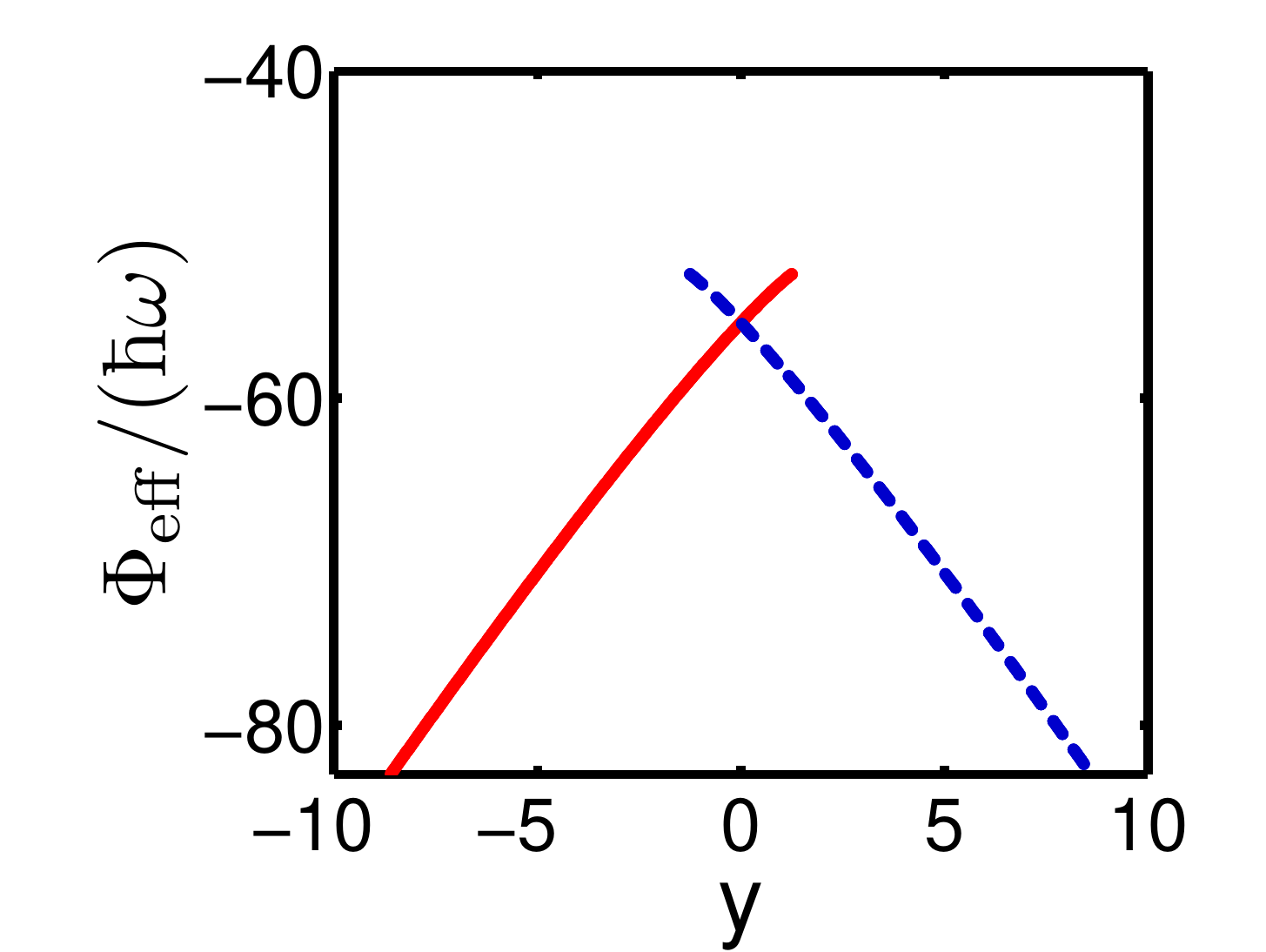}
   \label{Omega3_EGF:a}
 }
 
 \subfigure[]{
   \includegraphics[scale =0.26] {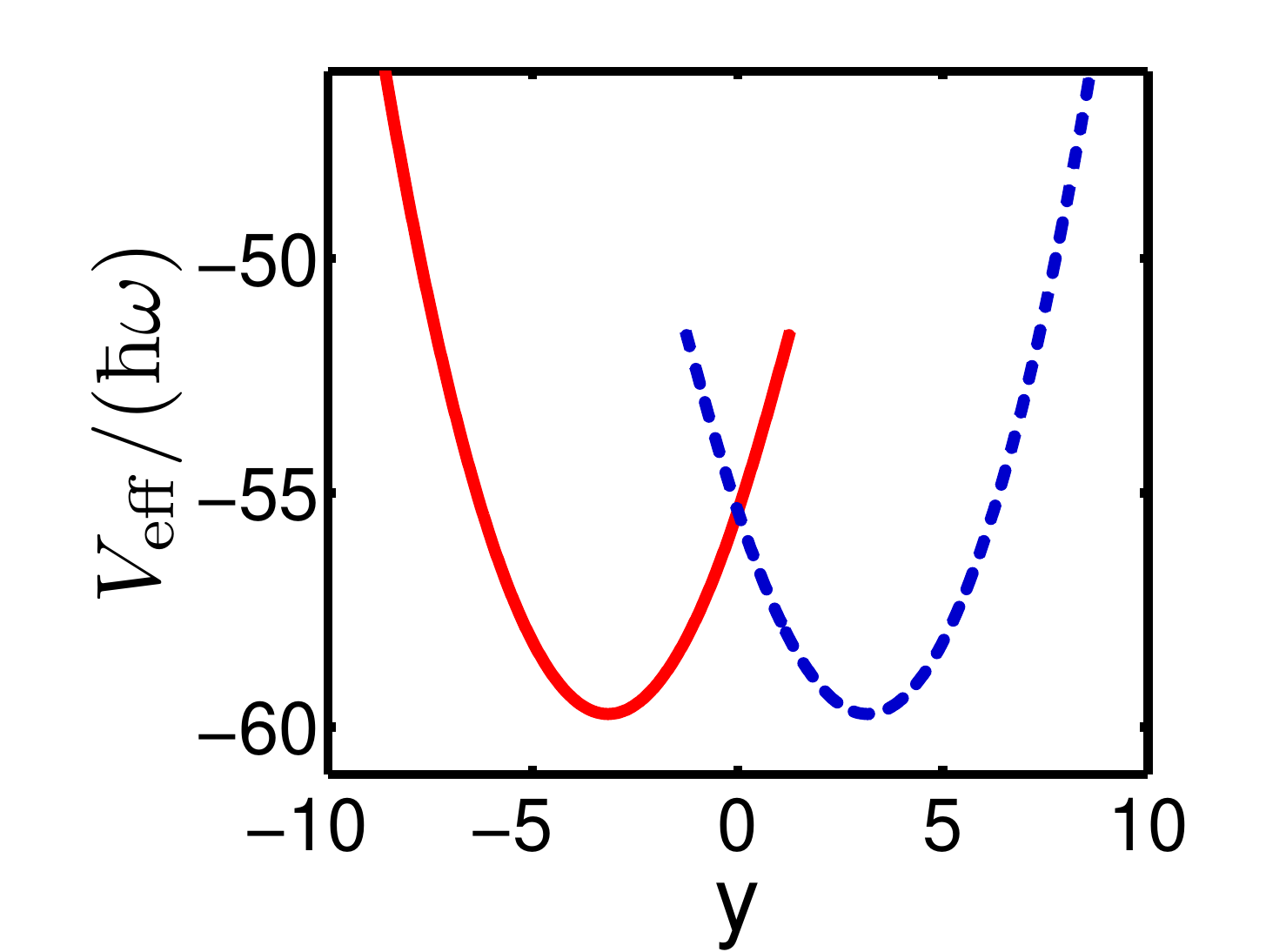}
   \label{Omega3_EGF:b}
 }} 

\mbox{ \subfigure[]{
   \includegraphics[scale =0.26] {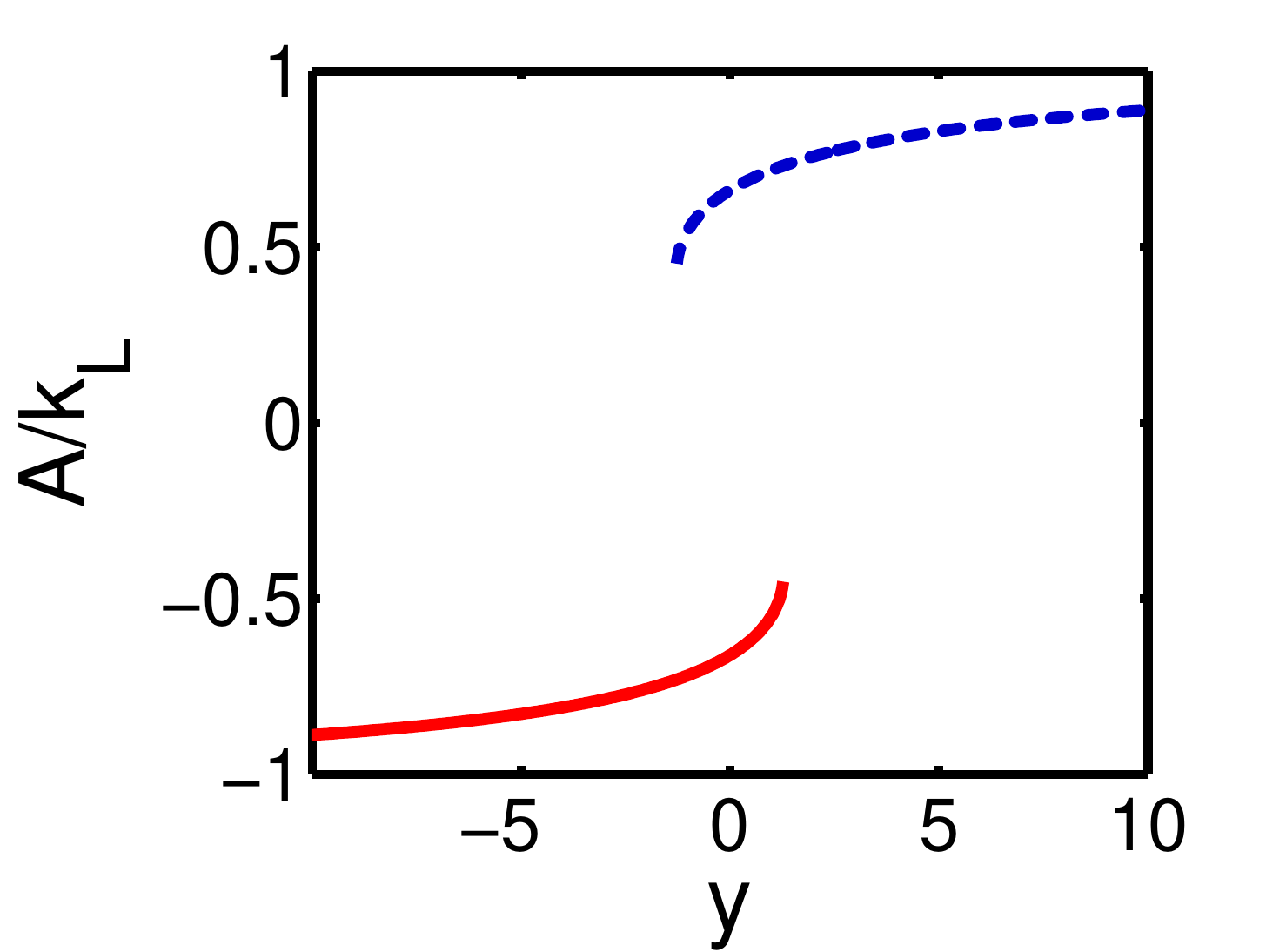}
   \label{Omega3_EGF:c}
 }

 \subfigure[]{
   \includegraphics[scale =0.26] {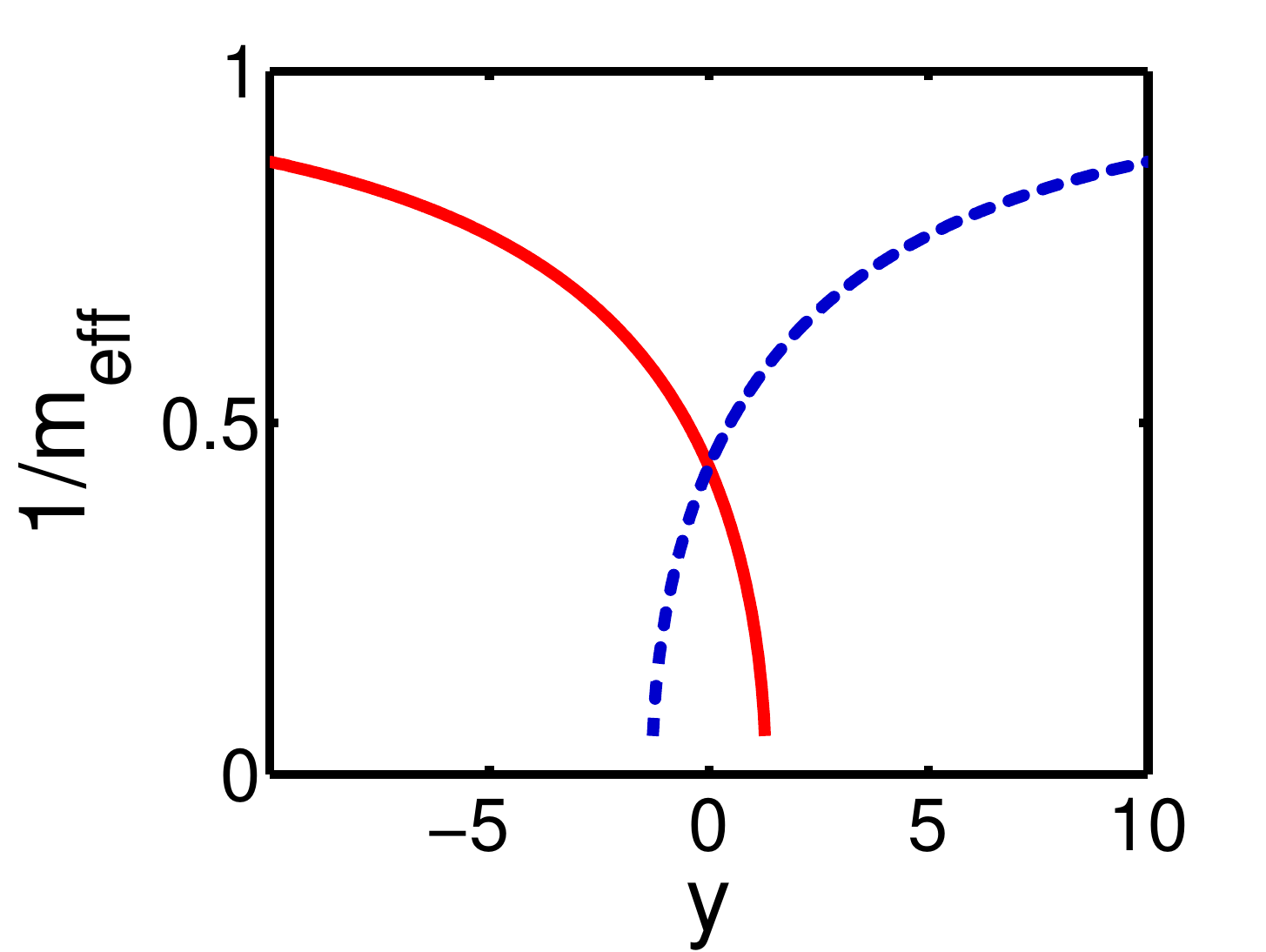}
   \label{Omega3_EGF:d}
 }}

\caption{(color online)
The figure shows the scalar potential $\Phi(y)$ (a), the effective trapping potential 
in y-direction $V_{\rm eff}(y)$ (b),
vector potential $A(y)$ (c) and inverse of the effective mass (d) for
$\Omega=3\ E_L$, $\beta=8\ \hbar \omega/a_0$ and $\gamma=1$. Values corresponding to the left minimum
of the spectrum are represented by a solid red line while the values corresponding to the left minimum
of the spectrum are represented by a dashed blue line (see the text for details).
}
\label{Omega3_EGF}
\end{figure}

To get the effective potential in $\hat{y}$ direction acting on left- and right-movers we define: 
$V_{\rm eff, L}(y)=\Phi_{L}(y) + \frac{1}{2}\gamma^2 y^2$, $V_{\rm eff, R}(y)=\Phi_{R}(y) + \frac{1}{2}\gamma^2 y^2$. 
In Fig. \ref{Omega3_EGF} we show
$\Phi_{L/R}(y)$, $V_{\rm eff, L/R}(y)$, $A_{L/R}(y)$ and $1/m_{\rm eff, L/R}(y)$ 
for $\Omega=3\ E_L$ and $\beta=8\ \hbar \omega/a_0$.
$V_{\rm eff, L/R}$ have minima at $y_{0,R/L}=\pm 3.2$ which explains the total 
density profile (Fig.~\ref{Omega3}(a)) which has maxima at $y= \pm 3.2$. 
  The position of two peaks in momentum distribution in Fig. \ref{Omega3}(c) can be understood as follows: for particles 
positioned near the minimum of $V_{\rm eff, L}$ in Fig. \ref{Omega3_EGF}(b), it is energetically favorable 
to have the $\hat{x}$-component of momentum approximately equal to $A(y_{0,L})$ and the $\hat{y}$-component near zero.  
Fig.~\ref{Omega3_EGF}(c) shows that $A(y_{0,L}) \approx -0.79\ k_L$, while from Fig. \ref{Omega3}
(c), we see that the momentum distribution is centered around $k_x=-0.80\ k_L$. The same explanation applies for the
momentum distribution of right-movers. 

\begin{figure}[ht]
\centering

\mbox{\subfigure[]{
   \includegraphics[scale =0.26] {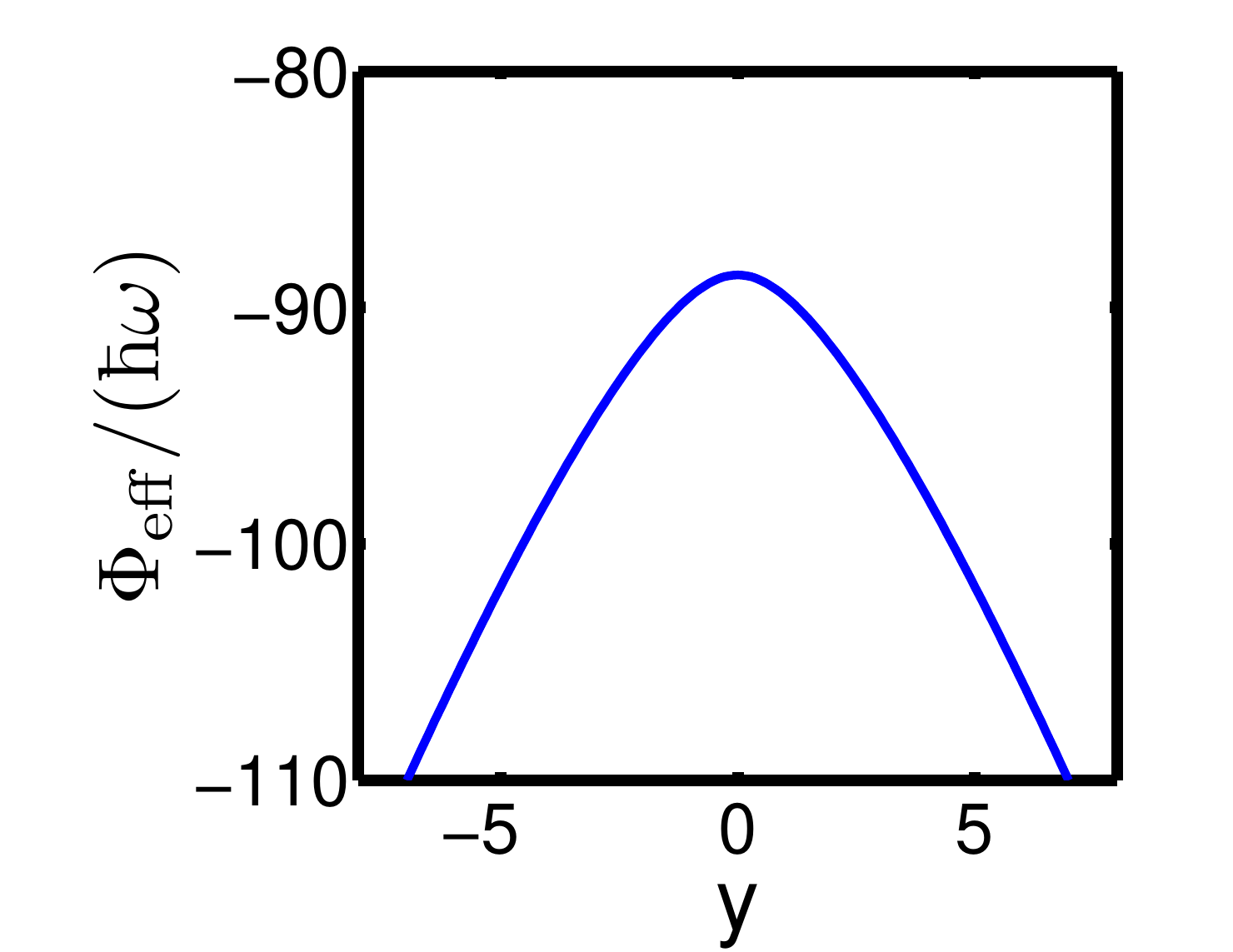}
   \label{Omega5_EGF:a}
 }
 
 \subfigure[]{
   \includegraphics[scale =0.26] {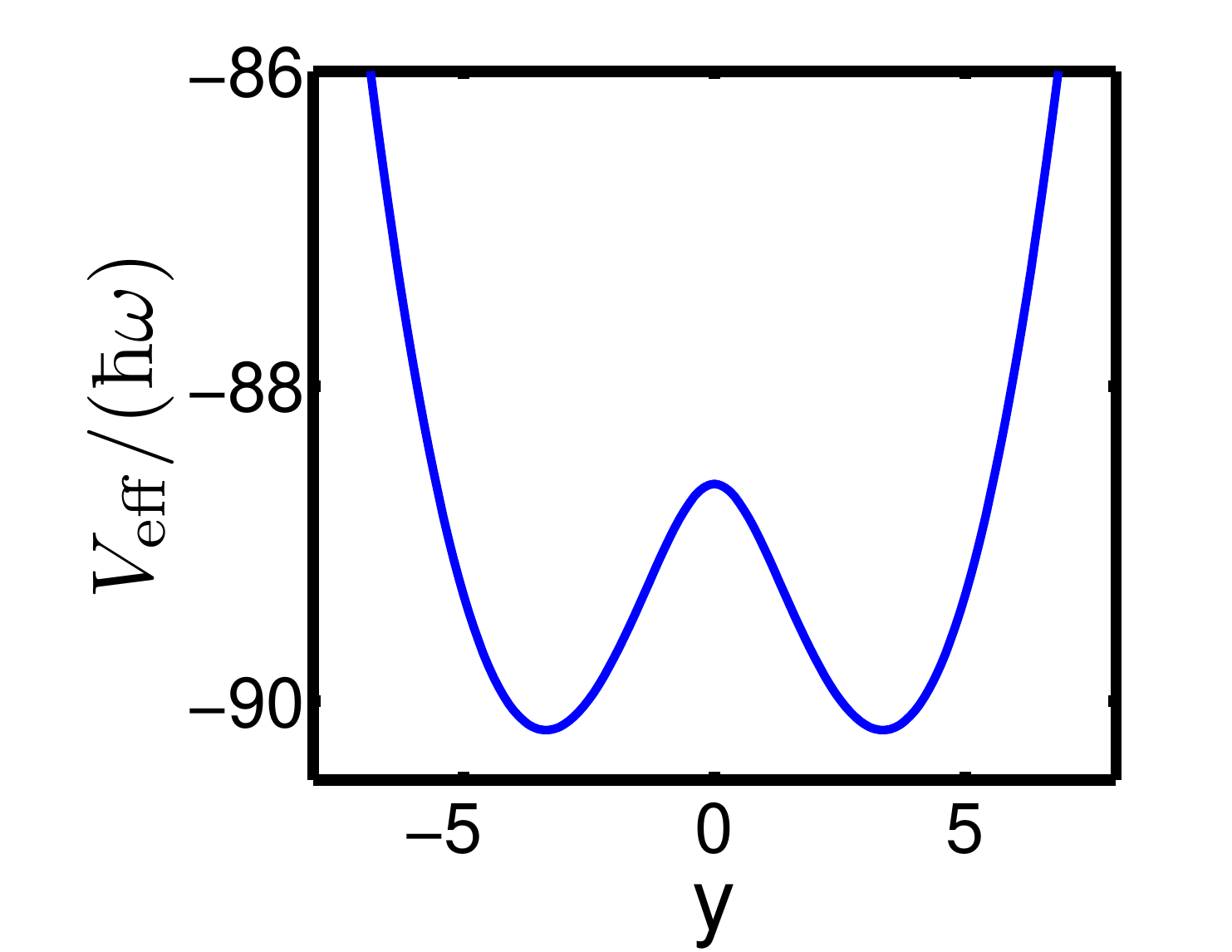}
   \label{Omega5_EGF:b}
 }} 

\mbox{ \subfigure[]{
   \includegraphics[scale =0.26] {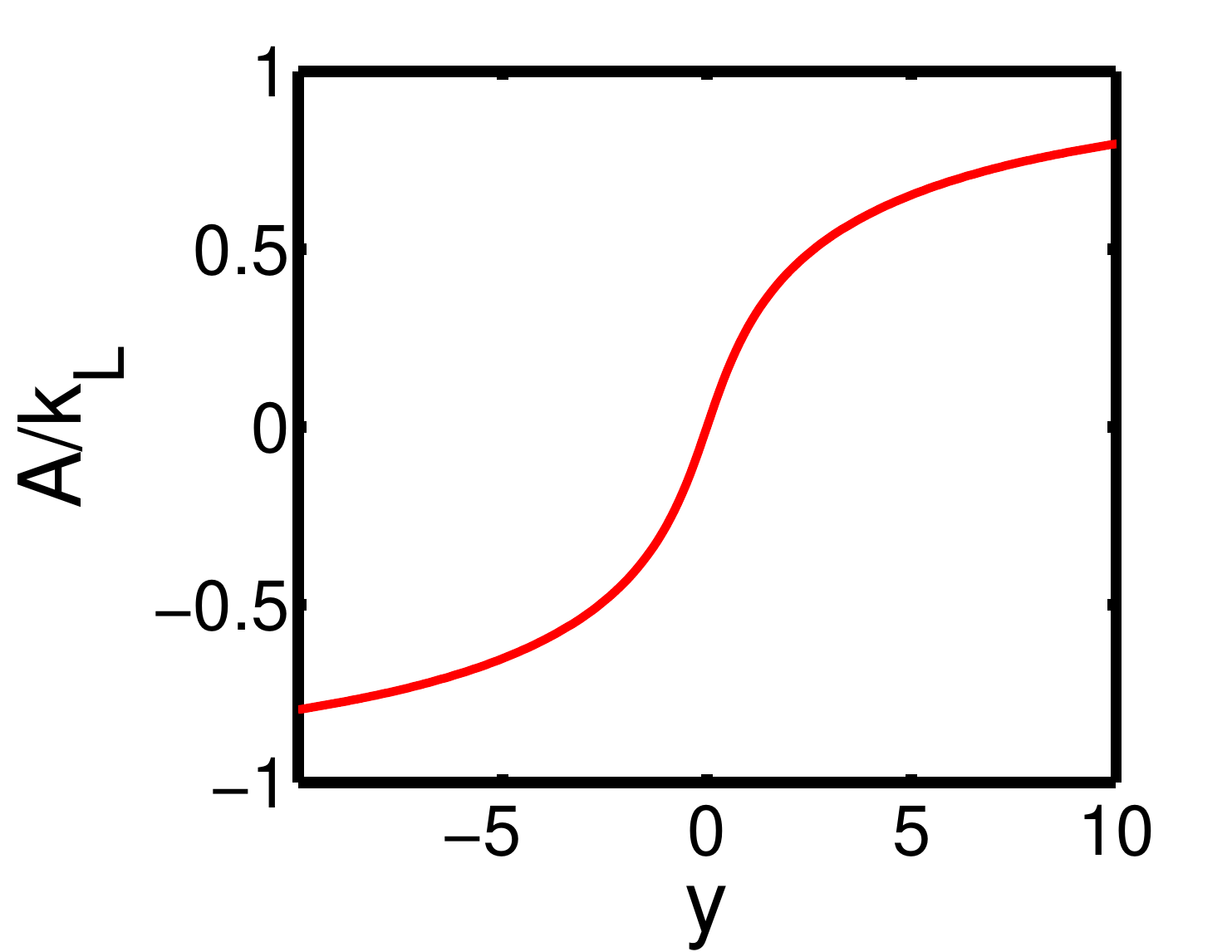}
   \label{Omega5_EGF:c}
 }

 \subfigure[]{
   \includegraphics[scale =0.28] {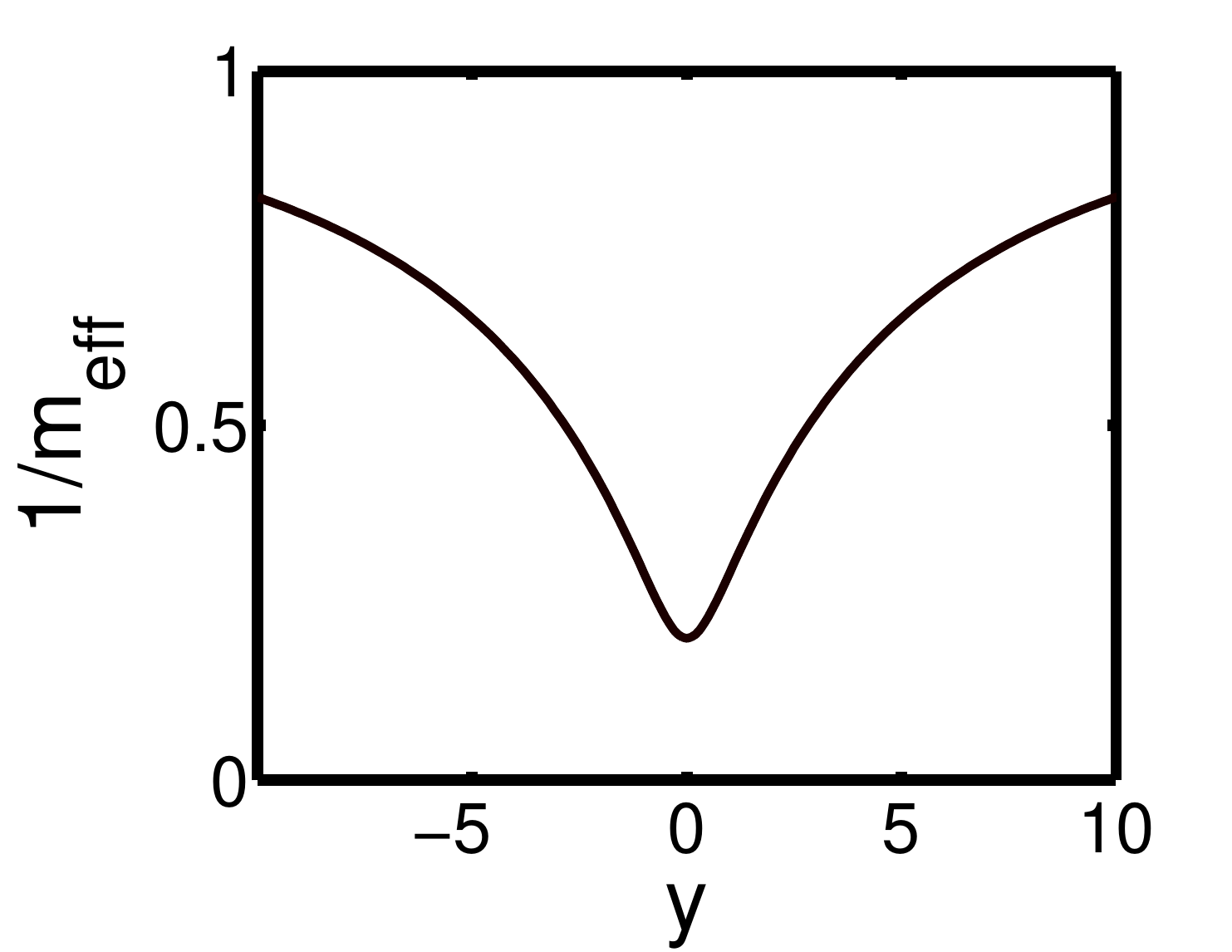}
   \label{Omega5_EGF:d}
 }}
\caption{(color online)
The figure shows the scalar potential $\Phi(y)$ (a), the effective trapping potential 
in y-direction $V_{\rm eff}(y)$ (b),
vector potential $A(y)$ (c) and inverse of the effective mass (d) for
$\Omega=5\ E_L$, $\beta=12\ \hbar \omega/a_0$ and $\gamma=1$. 
}
\label{Omega5_EGF}
\end{figure}

To investigate the regime with a single minimum in the spectrum ($\Omega \geq 4 E_L$)
we did calculations for parameters: $\Omega=5 E_L$, $\beta=12 \hbar \omega/a_0$ and $\gamma=1$ 
(Fig.~\ref{Omega5}). In this ``single-minimum'' case one might expect momentum distribution to be concentrated
around a single point as was observed in Ref.~\cite{Lin2011}. However, in spatially-dependent-detuning case this is
not necessarily true: the momentum distribution (Fig.~\ref{Omega5}(c)) shows two peaks around $k_x=\pm 0.55\ k_L$. Also,
the total density (Fig. \ref{Omega5}(a)) has a characteristic  
series of minima along $y=0$ line which come from vortices in the $\psi_{\uparrow}$ and $\psi_{\downarrow}$ 
wave-functions (Fig.~\ref{Omega5}(b)) created in the overlapping region of left- and right-movers.
The results can again be explained by the effective gauge field. The effective potential in $\hat{y}$-direction
$V_{\rm eff}(y)=\Phi(y)+\frac{1}{2}\gamma^2 y^2$ (Fig.~\ref{Omega5_EGF}(a)) has two minima at $y_{0,R/L}=\pm 3.4$ which explains
the density distribution which has maxima at $y=\pm 3.3$. Also, equation (\ref{H_prime}) tells us it is energetically 
favourable for particles near the left (right) minimum of $V_{\rm eff}(y)$ to have momentum around $A(y_{0,L})=-0.56\ k_L$ 
($A(y_{0,R})=0.56\ k_L$) (Fig.~\ref{Omega5_EGF}(c)) which explains momentum distribution.
We also note that in Fig.~\ref{Omega5_EGF}(c) $A(y)$ has a large gradient and therefore
magnetic field ($B_{\rm eff} \sim \partial A/ \partial y$) is strong around $y=0$ which may serve as an alternative explanation
of line of vortices appearing in Fig.~\ref{Omega5}(a).  

\begin{figure}[ht]
\centering

\subfigure[]{
   \includegraphics[scale =0.33] {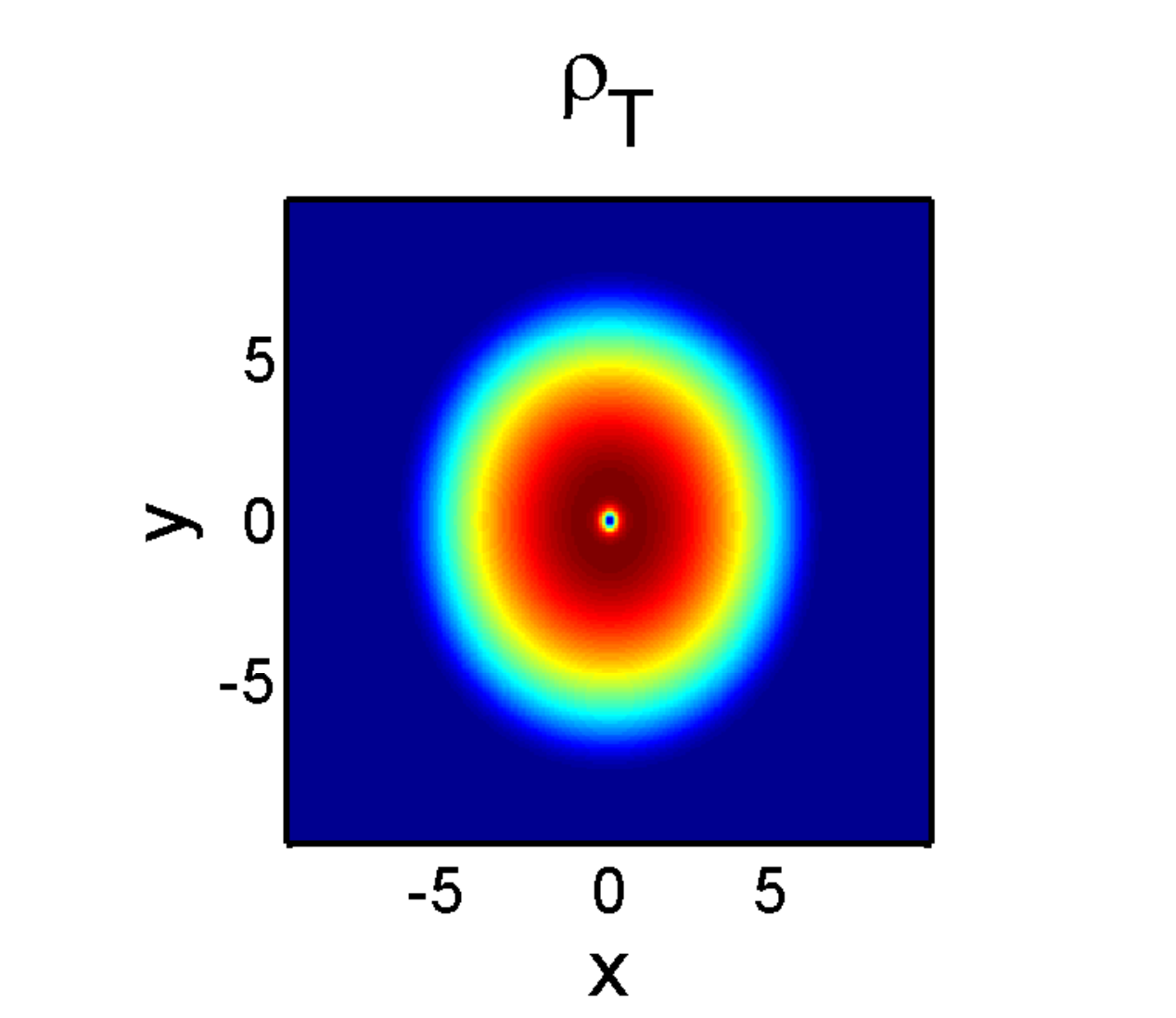}
   \label{Omega10:a}
 }
 
\subfigure[]{
   \includegraphics[scale =0.34] {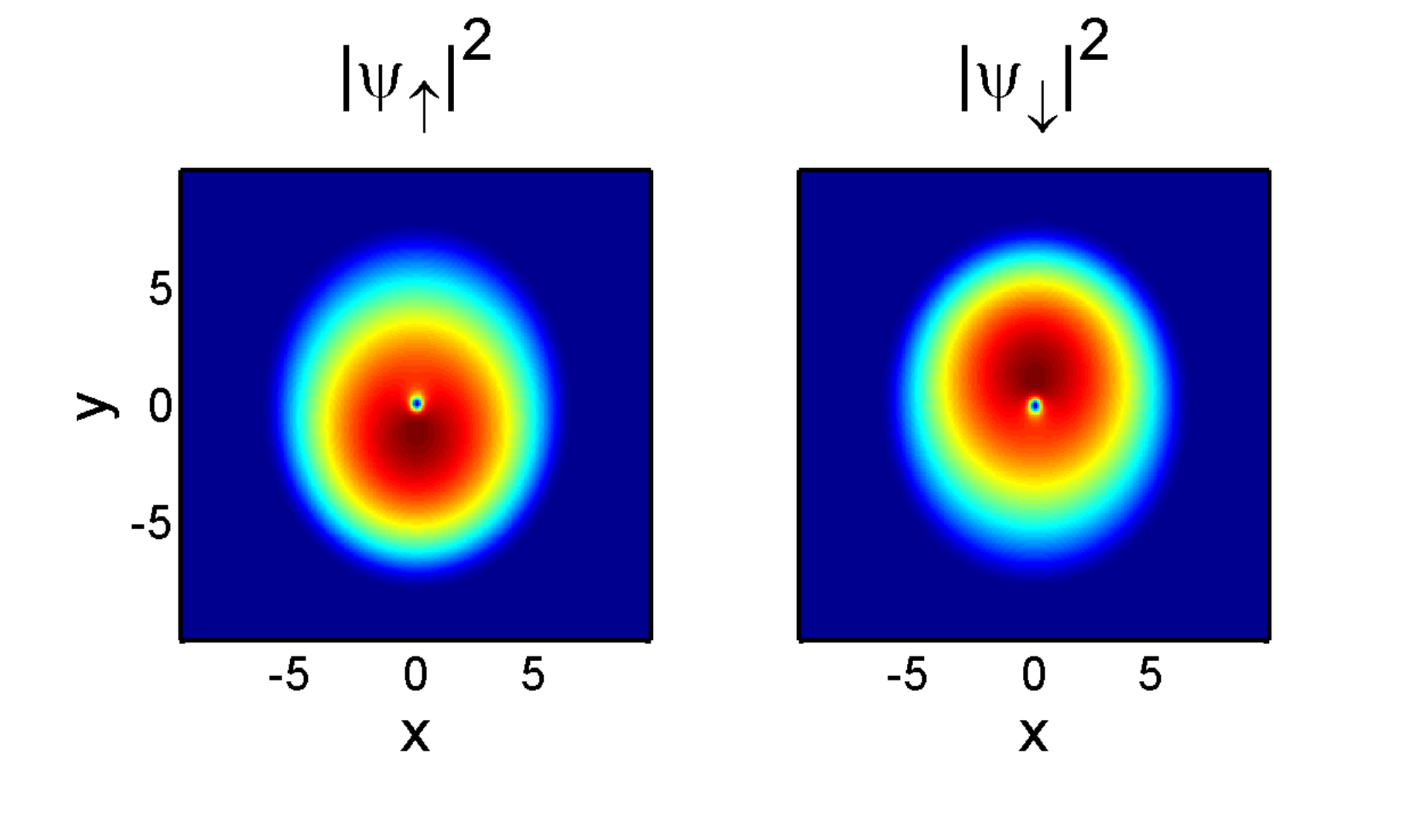}
   \label{Omega10:b}
 } 



\caption{(color online)
The figure shows results for $\Omega=10\ E_L$, $\beta=12\ \hbar \omega/a_0$
and $\gamma=1$. 
In (a) the total density is shown, while (b) and (c) show densities of 
$\lvert \uparrow \rangle$ and $\lvert \downarrow \rangle$ (c) components.
The vortex in the center appears for strong enough effective magnetic field. 
}
\label{Omega10}
\end{figure}

\begin{figure}[ht]
\centering

\mbox{ \subfigure[]{
   \includegraphics[scale =0.28] {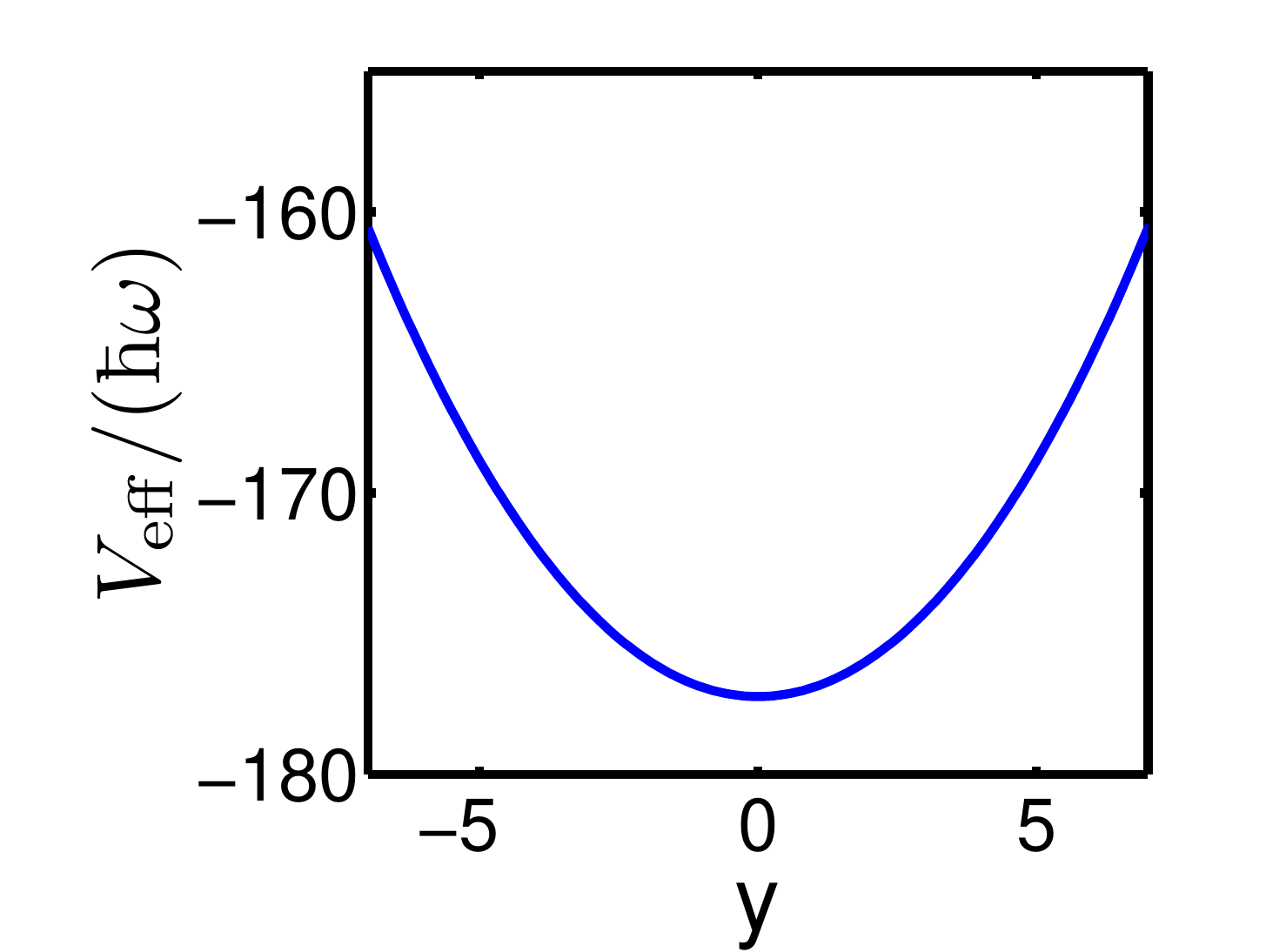}
   \label{Omega10_EGF:a}
 }

 \subfigure[]{
   \includegraphics[scale =0.28] {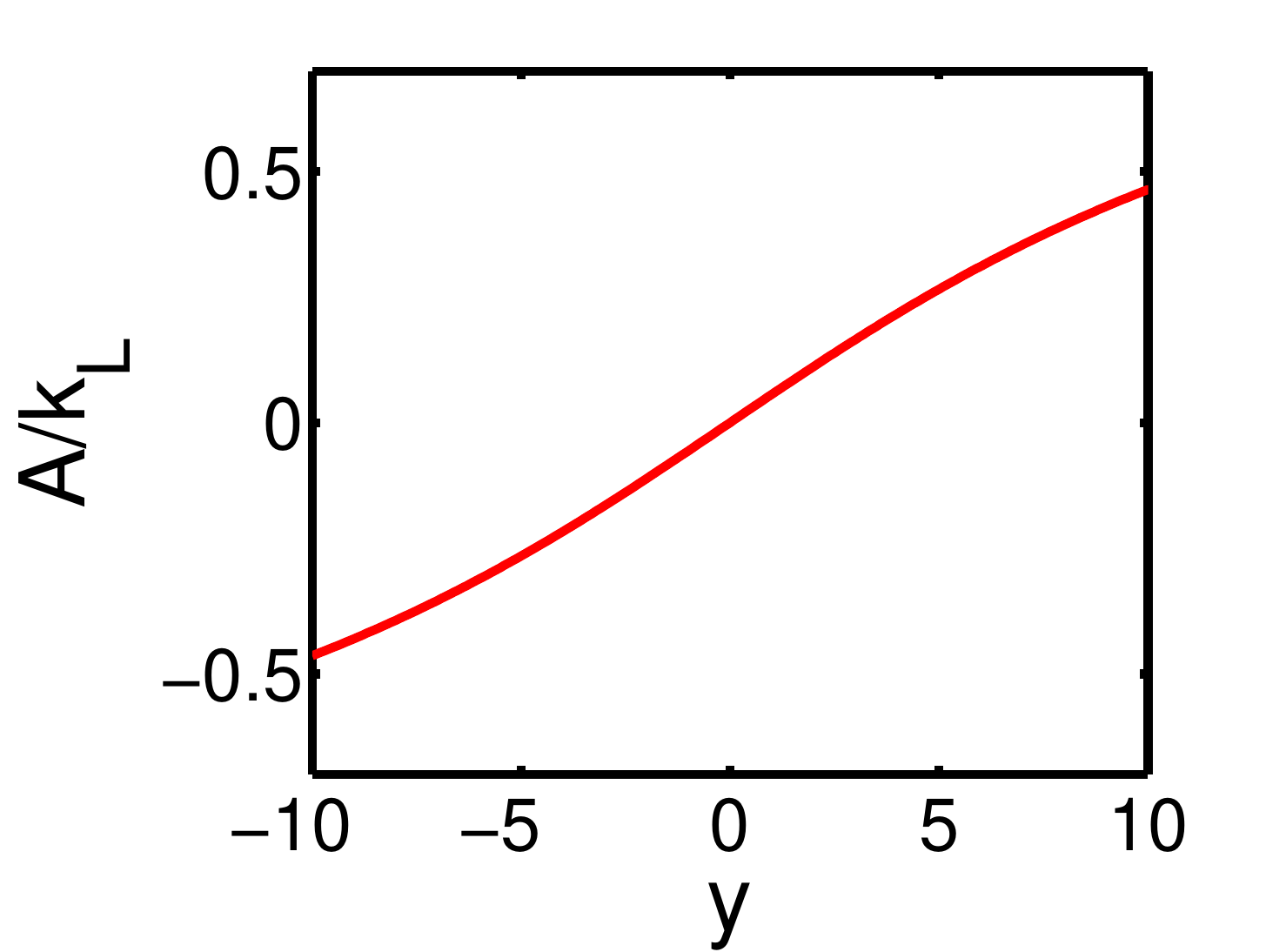}
   \label{Omega10_EGF:b}
 }}
 
\mbox{ \subfigure[]{
   \includegraphics[scale =0.28] {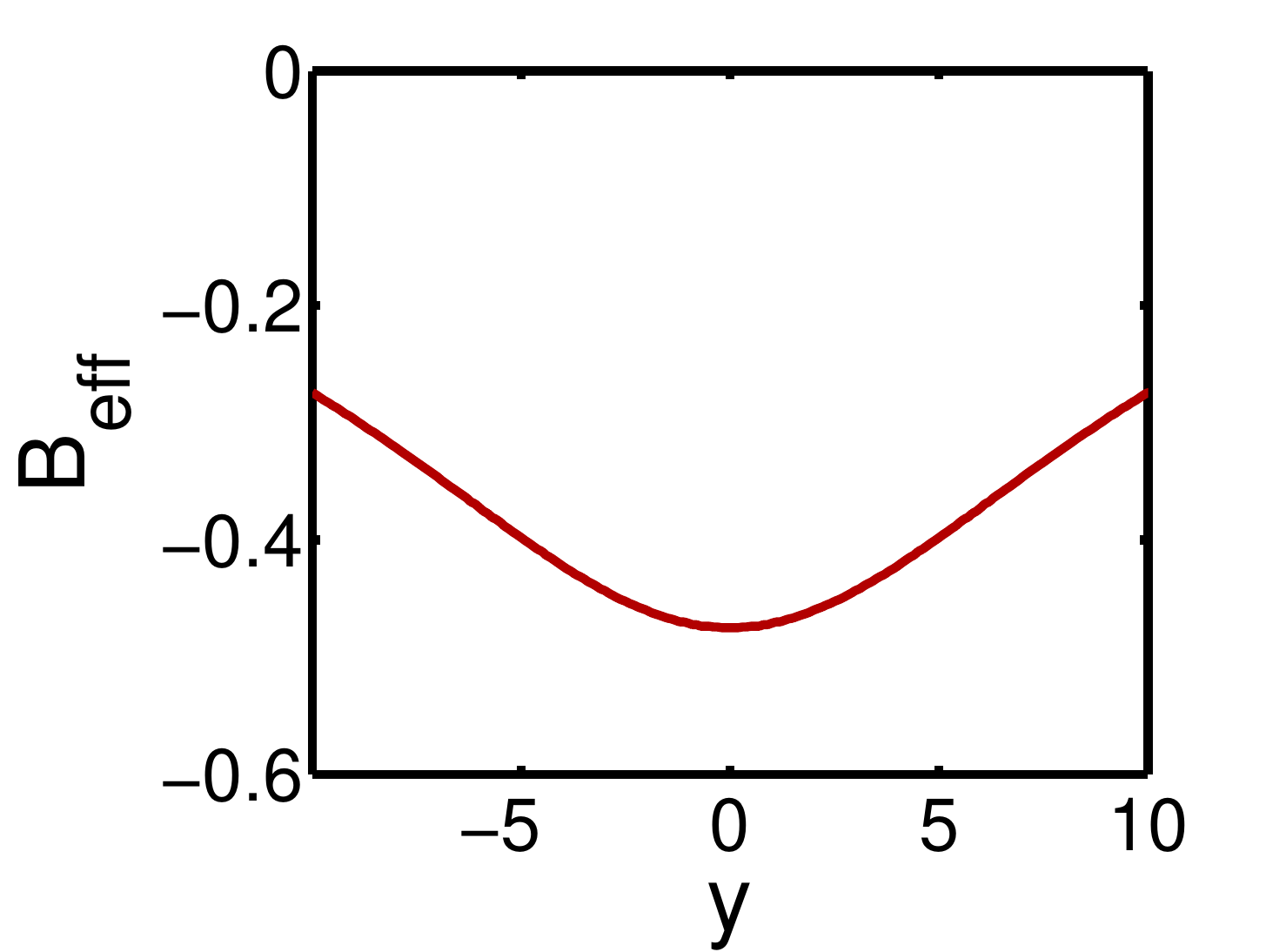}
   \label{Omega10_EGF:c}
 }

 \subfigure[]{
   \includegraphics[scale =0.28] {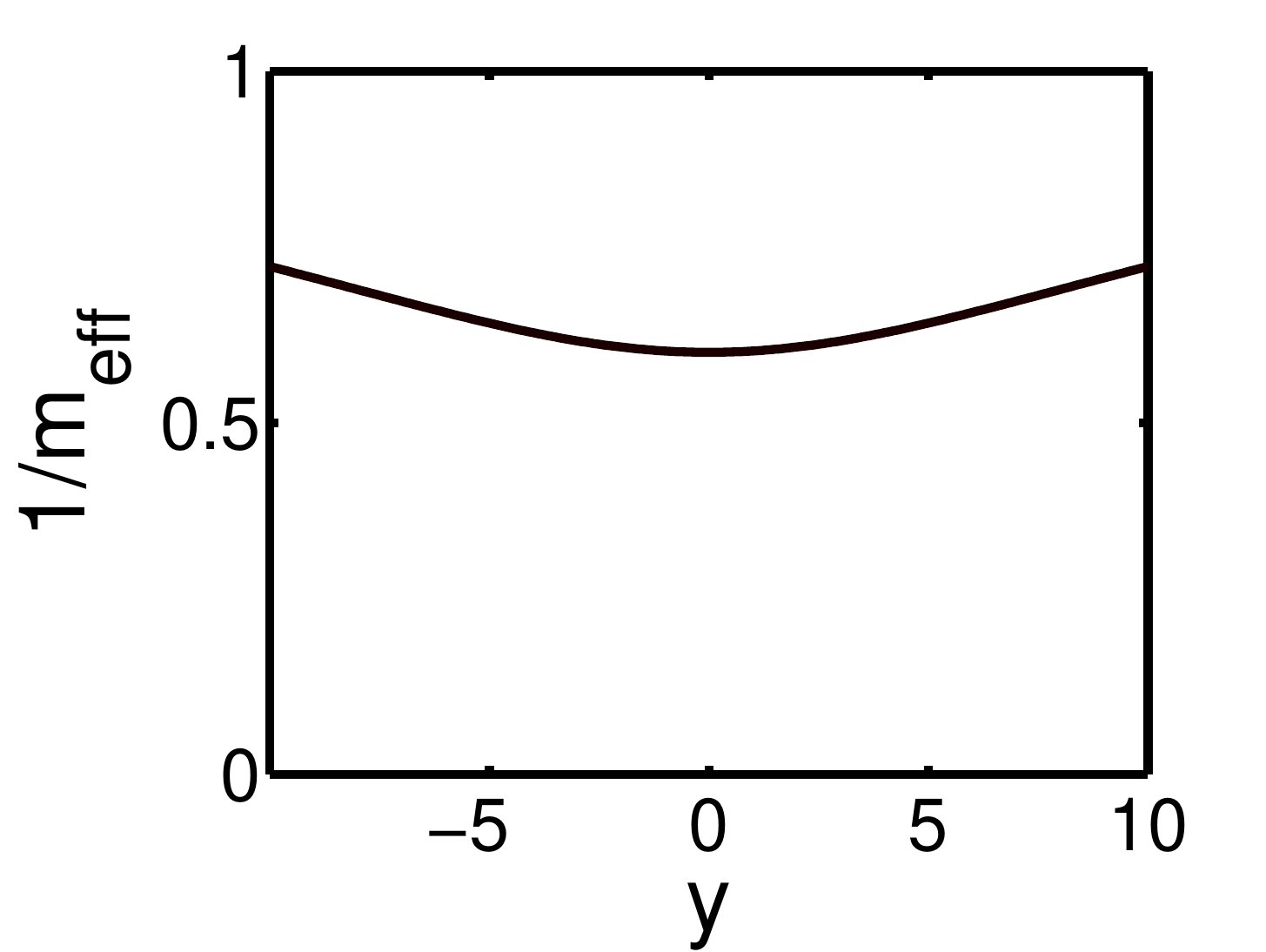}
   \label{Omega10_EGF:d}
 }}

\caption{(color online)
The figure shows the effective trapping potential in y-direction $V_{\rm eff}(y)$ (a),
vector potential $A(y)$ (b), the effective magnetic field $B_{\rm eff}$ (c) 
and inverse of the effective mass (d) for $\Omega=10\ E_L$, 
$\beta=12\ \hbar \omega/a_0$ and $\gamma=1$.
}
\label{Omega10_EGF}
\end{figure}

We now study the system with strong Raman coupling $\Omega$ and weak detuning gradient $\beta$ (i.e. $\beta$
is not large enough to produce spatial separation of a cloud along $\hat{y}$ as in previous cases). 
  Results for $\Omega=10\ E_L$, $\beta=12\ \hbar \omega/a_0$ are shown in Fig.~\ref{Omega10} and can be explained
by the associated effective gauge field shown in Fig.~\ref{Omega10_EGF}. The total density (Fig.~\ref{Omega10}(a)) and
$|\psi_{\uparrow}|^2$, $|\psi_{\downarrow}|^2$ (Fig.~\ref{Omega10}(b)) show the existence of a vortex
in the centre of the cloud. The vortex appears only for strong enough effective magnetic 
field which is tuned by changing $\beta$. We define the effective magnetic field ${\mathbf B}_{\rm eff}=\bm {\nabla} 
\times {\mathbf A}(y)$ and in our case, (${\mathbf A}=(A(y),0,0)$),  ${\mathbf B}_{\rm eff}=-\frac{\partial
A(y)}{\partial y} \hat{z}$. The magnetic field points in the $\hat{z}$ direction, depends on
$y$, and is constant along $x$. We also note that since $m_{\rm eff}(y) \neq 1$ 
(Fig. \ref{Omega10_EGF}(d)), the 
effective equations will differ from those for an ordinary charged particle in a magnetic 
field $B_{\rm eff}(y) \hat{z}$. The vector potential $A(y)$ and the effective magnetic field ${\mathbf B}_{\rm eff}(y)$ are shown in Fig.~\ref{Omega10_EGF}(b,c). 

It is useful to know the
critical field needed for vortex creation and we may get a crude estimate by using the equation for critical magnetic 
field of a single-component 2D gas in the Thomas-Fermi limit: 
$B_c=4 (a_0/R)^2 \ln \big( 0.888(R/a_0)^2 \big)$, where $R$ is the Thomas-Fermi radius of the cloud 
\cite{Pethick2008}. We take 
$R=6.5\ a_0$ (the size of our cloud), which gives $B_c \approx 0.35$. It is important to notice that larger number of particles 
or stronger interactions increase $R$, which lowers the critical field ($B_c$ decreases with 
increasing $R$). To find $B_c$, we did simulations for $\Omega=10\ E_L$, $\gamma=1$ and 
for different values of $\beta$ (which controls the strength of the effective magnetic field). 
We found that the vortices start to appear for a critical effective magnetic field $B_c \approx 0.34$,
which is very close to our estimate presented above.

\begin{figure}
\centerline{
\mbox{\includegraphics[width=0.36\textwidth]{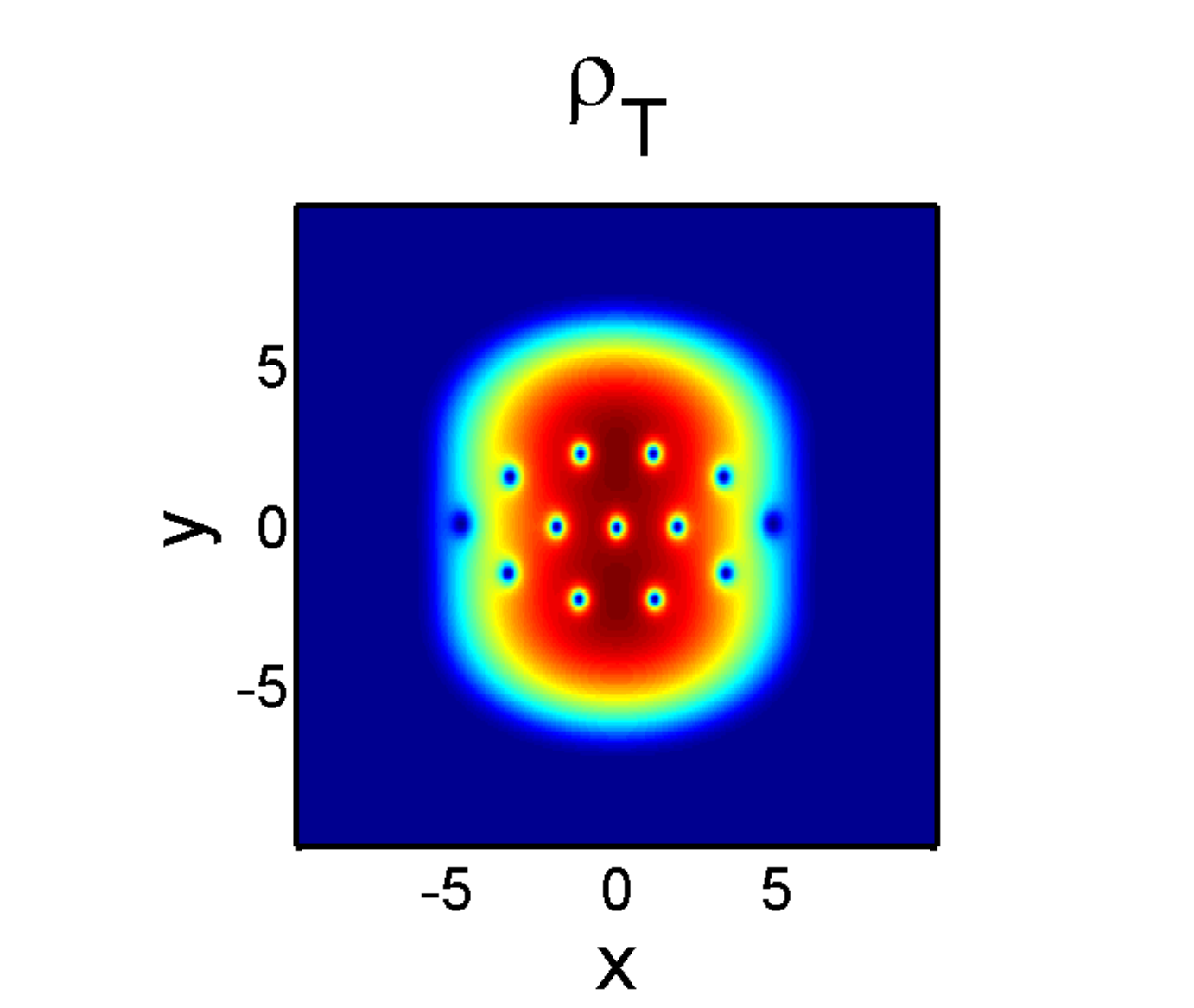}}
}
\caption{(color online)
The figure shows the total density for $\Omega=10\ E_L$, $\beta=40\ \hbar \omega/a_0$
and $\gamma=1.85$.  
}
\label{vortex_lattice}
\end{figure}

  If the effective field is strong enough, a vortex ``lattice'' is formed, as shown in Fig. \ref{vortex_lattice},
  which corresponds to $\Omega=10\ E_L$, $\beta=40\ \hbar \omega/a_0$ and $\gamma=1.85$.
From the figure, we see that vortices are concentrated along the $x$-axis and around $y=0$. This is because 
$B_{\rm eff}(y)$ is not homogeneous, i.e. the field is strongest at $y=0$ and it weakens with
increasing $|y|$. We had to increase trapping strength in the $\hat{y}$ direction ($\gamma=1.85$)
because scalar potential $\Phi(y)$ separates the clouds (e.g. see Fig. \ref{Omega5_EGF}(a))
and for a weaker trapping strength, the effective potential would have two minima (it would look like effective potential in 
Fig. \ref{Omega5_EGF}(b)).

  The most interesting regime is the one in which left and right moving phases 
$\big( \psi_{\uparrow L}({\mathbf r})$, $\psi_{\uparrow L}({\mathbf r}) \big)$ and 
$\big( \psi_{\uparrow R}({\mathbf r})$, $\psi_{\uparrow R}({\mathbf r}) \big)$ are spatially 
separated along $\hat{y}$ direction and there is a vortex (or vortices) in each phase in 
addition to a vortex line. This requires double minimum structure of the effective
potential in $\hat{y}$ direction $V_{\rm eff}(y)$, which separates the phases and strong enough
effective magnetic field in each phase to create additional vortices, which tend to appear in pairs (i.e. the number
of vortices is equal in both phases which is a consequence of the fact that in our simulations
the effective gauge field is symmetric with respect to reflection about $y=0$ line and interactions are almost
spin-independent).
  
\begin{figure}[ht]
\centering

\centerline{
\mbox{\includegraphics[width=1.0\columnwidth]{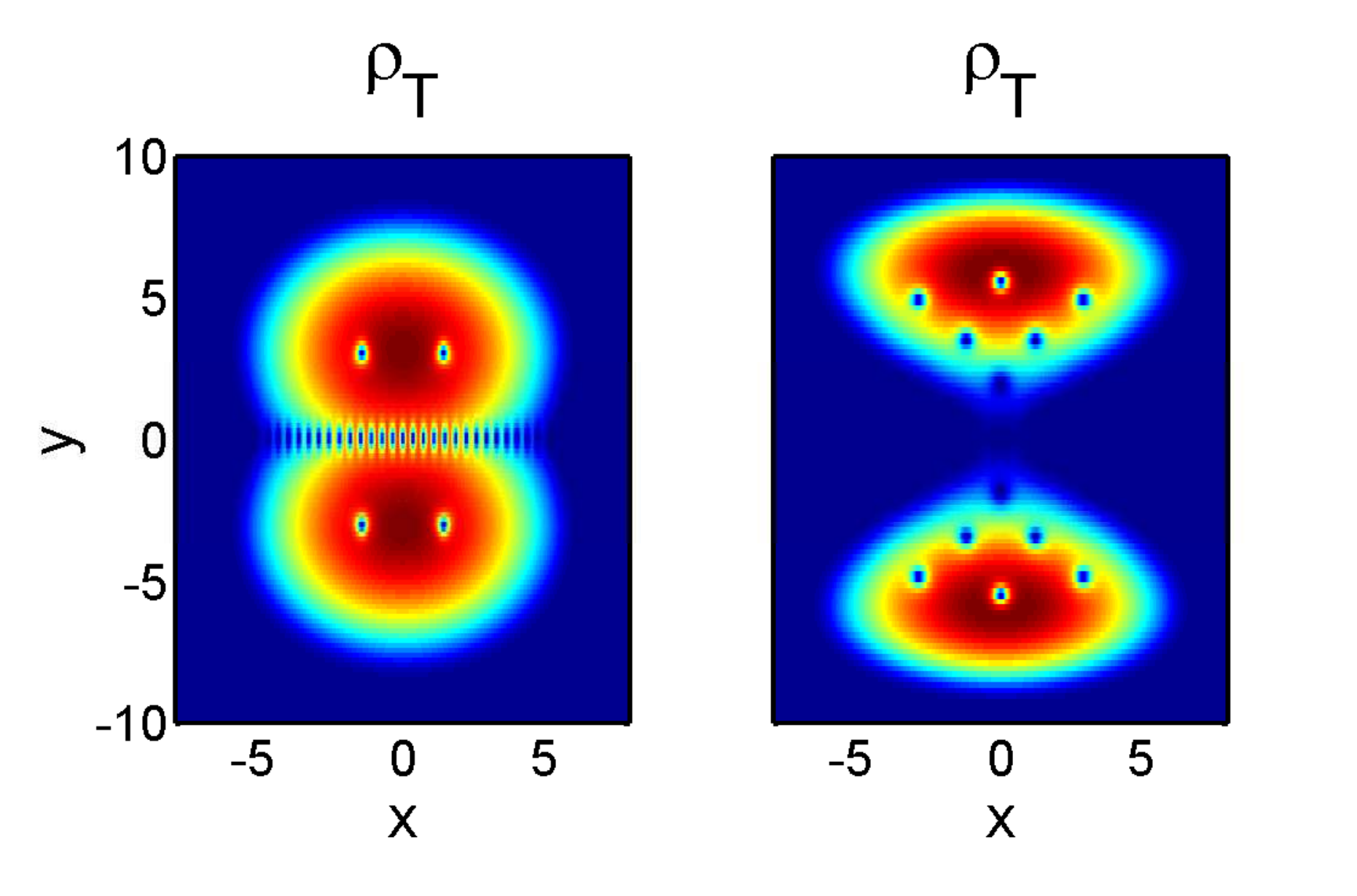}}
\put(-178,0){(a)}
\put(-70,0){(b)}
}

\caption{(color online)
Figures show separated left and right phases with
vortices in each phase. Trapping frequency is $\omega=2\pi \times 10\ \rm Hz$.
 Figure (a) shows total density for $\Omega=4\ E_L$, 
$\beta=20\ \hbar \omega/a_0$ and $\gamma=1.3$.  
Figure (b) shows total density for $\Omega=10\ E_L$, $\beta=150\ \hbar \omega/a_0$ 
and $\gamma=2.75$.  
}
\label{double_vortex}
\end{figure}

 In Fig. \ref{double_vortex}(a), we show results for $\omega=2\pi \times 10\ \rm Hz$, 
$\Omega=4\ E_L$, 
$\beta=20\ \hbar \omega/a_0$ ($a_0=\sqrt{\hbar/({m \omega})}$). By choosing $\omega=2\pi \times 10\ \rm Hz$, parameter
$k_L^{\prime}$ in dimensionless GP equations (\ref{dimensionless_GP2}) becomes $k_L^{\prime}=18.83$, while interaction
coefficients stay the same ($g_1=1600$, $g_2=1593$ and $g_{12}=1593$). Having larger $k_L^{\prime}$ means we can
create stronger effective magnetic field.
We increased the
trapping frequency in the $y$-direction ($\gamma=1.3$) to bring two phases closer to 
$y=0$, where the effective field is stronger (to counter the effective scalar potential $\Phi(y)$, which separates the phases). 
  In Fig. \ref{double_vortex}(b) we show results for $\omega=2\pi \times 10\ \rm Hz$, 
$\Omega=10\ E_L$, $\beta=150\ \hbar \omega/a_0$ and $\gamma=2.75$. Here the left and right
phases are completely separated in space and the effective magnetic field is strong enough
to produce multiple vortices in each phase. Also, it is clear that the vortices are not located in centers 
of two phases, but are positioned closer to $y=0$ which is expected because the field is stronger near
$y=0$. 

  It is important to discuss the means of experimentally observing results we presented.
We concentrate on the time-of-flight imaging, which is widely used  to
probe cold-atoms systems. The time-of-flight picture here will be determined by the underlying 
momentum distribution of particles. If  this momentum distribution consists of two separated peaks, the
 initial cloud will strongly separate during expansion (see for example \cite{Lin2009,Lin2011}). 
  We note that due to the transformation $\psi^{\prime}_{\uparrow}({\mathbf r})=
\psi_{\uparrow}({\mathbf r})e^{-ik_L x}$, $\psi^{\prime}_{\downarrow}({\mathbf r})=
\psi_{\downarrow}({\mathbf r})e^{ik_L x}$ used
when deriving Hamiltonian (\ref{H_detuning2}), the real momentum distribution of $\lvert \uparrow \rangle$ particles will 
 in fact be shifted by $k_L$ with respect to the momentum distribution shown in figures and the momentum distribution of 
$\lvert \downarrow \rangle$ particles is shifted by $-k_L$ (see \cite{Lin2011}). 
  In the case of $\Omega=0$, 
both $\lvert \uparrow \rangle$ and $\lvert \downarrow \rangle$ particles will have zero
average momentum, which means both components of the condensate
will expand, while the position of centre of mass will be stationary  during time-of-flight.
For $\Omega=3\ E_L$ and $\beta=8\ \hbar \omega/a_0$, we expect four separated clouds to be seen 
in the time-of-flight:  since the real momentum distributions of $\lvert \uparrow \rangle$ and $\lvert \downarrow \rangle$ 
particles are shifted by $k_L$ and $-k_L$, there will be two clouds of $\lvert \uparrow \rangle$ 
particles with average momenta of $0.2\ k_L$ (larger cloud) and $1.8\ k_L$ (smaller cloud) 
and two clouds of $\lvert \downarrow \rangle$ particles with average momenta of $-0.2\ k_L$ 
(larger cloud) and $-1.8\ k_L$ (smaller cloud). It is important to notice that the vortex line
will not be easily visible in those images, because it exists only due to the the overlap of  the wave-packets with different average momenta. During the time-of-flight, two wave-packets $\big( \psi_{\uparrow,L}, \psi_{\uparrow,R} \big)$
or $\big( \psi_{\downarrow,L}, \psi_{\downarrow,R} \big)$ separate, which means that they do
not overlap any more and there is no clear vortex line present.  For the case in Fig.~\ref{double_vortex}, the
vortices in each phase will be visible since they are not a result of overlapping the left- and right-moving condensates.

\section{Conclusion}
In this paper, we have investigated realistic experimental methods that can be used to create vortex excitations in spin-orbit-coupled Bose-Einstein condensates. The main conclusion of the work is that due to a complicated interplay between effects associated with the applied laser fields and rotation, the resulting state of the spin-orbit BEC under additional perturbations is highly non-universal and depends strongly on the system parameters and specific laser schemes. In particular, we argued that a spin-orbit BEC under rotation of the trap alone does not achieve a thermodynamically stable state at all, but acquires a complicated non-equilibrium dynamics that eventually leads to heating and the destruction of the condensate.

We have also suggested two alternative experimental methods to mimic an Abelian ``orbital'' magnetic field that involve 
either rotation of the entire experimental setup, or a spatially-dependent detuning. We performed numerical simulations of the resulting thermodynamically stable density distributions, focusing mostly on the M-scheme that has already been realized  experimentally. This scheme gives rise to an ``Abelian'' spin-orbit-coupling with a well-understood ground state that we used as a basis of our numerical simulations that showed topological excitations above the ground state. We expect that the predicted vortex configurations, in particular vortices appearing in pairs in the spatially-separated left- and right-moving regions, would be straightforward to observe experimentally, as all necessary ingredients are already experimentally available.

Finally, we would like to mention that to this point only an ``Abelian'' spin-orbit-coupling scheme has been actually realized in experiment \cite{Lin2011} and we mostly focused here on vortex topological excitations in such systems with a well-understood ground state. What remains of great interest of course is an expermiental realization of a truly ``non-Abelian'' spin-orbit interaction (either of pure Rashba or Dresselhaus type or a non-equal mixture of those), which can be achieved using laser schemes described in Secs.~\ref{Sec:different_schemes}B, \ref{Sec:different_schemes}C, and Refs.~\cite{Stanescu2007,Campbell2011}. Note that it was argued theoretically~\cite{Stanescu2008} that in the Rashba-Dresselhaus system with single-particle dispersion of the double-well type, a fragmented condensed state \cite{Ashhab2003,Baym2006} can be selected by energetics for repulsive interactions that do not break the underlynig Kramers symmetry. This state arises because repulsive interactions in the real space tend to localize particles in the dual momentum space per the fundamental Heisenberg uncertainty principle. This robust argument together with the protection provided by Kramers symmetry and momentum conservation (modulo finite-size effects due to the trap) suggest that the long-sought-after fragmented BEC (which takes the form of a many-body Schr{\"o}dinger's cat-state in this case~\cite{Stanescu2008}) is more stable in spin-orbit-coupled systems than that in BECs confined to real-space double-well potentials and hence can be observed experimentally. Topological excitations above this exotic ground state are expected to also be of exotic nature and may potentially realize much of the exciting physics discussed in the context of multi-component superconductors \cite{Babaev2004,Babaev2007}. Finally, the nature of the ground state and topological excitations above it in the pure bosonic Rashba model remain of great interest as well. Depending on the interaction parameters, this model with a continuous ring of minima on a circle in momentum space, may potentially host topological BECs, spontaneous symmetry-broken states 
\cite{Gopalakrishnan2011}, and exotic Bose-liquid states~\cite{Varney2011}, where strong quantum fluctuations prohibit order even at zero temperature.

\acknowledgments

The authors are grateful to Austen Lamacraft, Egor Babaev, and Sankar Das Sarma for useful discussions.
This work was supported by AROÕs atomtronics MURI (JR and IBS) and US-ARO grant, ``Spin-orbit-coupled BECs '' (TAS and VG). 

{\it Note added --}      After this work was completed, we became aware of two papers which study spin-orbit-coupled
BECs under rotation [\onlinecite{XXu2011,Zhou2011}]. The fundamental assumptions in these papers are qualitatively 
different from our theory, in that Refs.~[\onlinecite{XXu2011,Zhou2011}] start with an effective spin-orbit-coupled Hamiltonian and assume that it remains stationary under rotation. This is in contrast to our theory, where we consider realistic experimental schemes, where rotation is shown to lead to a different description.

%

\begin{appendix}

\section{M-scheme with rotating trap and spin-orbit lasers}
\label{app:scheme1_a}

The Hamiltonian in the rotating frame (\ref{eq:H_RF}) is:
\begin{equation}
\begin{split}
\hat{H}_{RF} & = \left[ \frac{\hbar^2 \hat{{\mathbf k}}^2}{2m} + V({\mathbf r}) 
- \omega_r \hat{L}_z \right] \check{1} \\
& + \begin{pmatrix}
\hbar \left( -\omega_z + \omega_q \right)& 0& 0 \\
0& 0& 0 \\
0& 0& {\hbar \omega_z} 
\end{pmatrix} \\ 
& + \sqrt{2}\Omega \check{\sigma}_{3,x} \cos(2 k_L x + \Delta \omega_L t) 
- \omega_r \check{\sigma}_{3,z}.
\end{split}
\label{app:scheme1_RF_3}
\end{equation}
The Hamiltonian becomes time-independent if we transfer to the rotating-wave frame and if we do 
the rotating-wave approximation:
\begin{equation}
\begin{split}
&\hat{H}_{RF} = \left[ \frac{\hbar^2 \hat{{\mathbf k}}^2}{2m} + V({\mathbf r}) 
-\omega_r \hat{L}_z \right] \check{1} +
\begin{pmatrix}
{\delta + \hbar \omega_q}& 0& 0 \\
0& 0& 0 \\
0& 0& -\delta 
\end{pmatrix} \\
& \quad + \frac{\Omega}{\sqrt{2}} \check{\sigma}_{3,x} \cos(2 k_L x) 
- \frac{\Omega}{\sqrt{2}} \check{\sigma}_{3,y} \sin(2 k_L x)
- \hbar \omega_r \check{\sigma}_{3,z},
\end{split}
\end{equation}
where $\delta=\hbar (\Delta \omega_L - \omega_z)$.
We set quadratic Zeeman shift $\hbar \omega_q$ to be much greater than $\Omega$ and $\delta$
so we may restrict to the subspace spanned by 
$\lbrace \lvert m_z=0 \rangle, \lvert m_z=-1 \rangle \rbrace$:
\begin{equation}
\begin{split}
\hat{H}_{RF,2} &=\left[ \frac{\hbar^2 \hat{{\mathbf k}}^2}{2m} + V({\mathbf r})
- \omega_{r} \hat{L}_z \right] \check{1} 
+ \frac{\Omega}{2} \check{\sigma}_x \cos(2 k_L x) \\
& - \frac{\Omega}{2} \check{\sigma}_y \sin(2 k_L x) 
+
\begin{pmatrix}
0& 0\\
0& \hbar \omega_r - \delta
\end{pmatrix},
\end{split}
\label{app:scheme1_RF_2}
\end{equation}
where $\check{1}$ is 2 $\times$ 2 unit matrix and $\check{\sigma}_{x,y,z}$ are 2 $\times$ 2
Pauli matrices.
Since there are effecively two internal degrees of freedom we introduce pseudospin-1/2 notation, 
i.e. we define $\lvert \uparrow \rangle \equiv \lvert m_z=0 \rangle$, 
$\lvert \downarrow \rangle \equiv \lvert m_z=-1 \rangle$.
  We follow the steps in \cite{Lin2011} and make transformation: 
$\psi^{\prime}_{\uparrow}({\mathbf r})=\psi_{\uparrow}({\mathbf r})e^{-ik_L x}$,
$\psi^{\prime}_{\downarrow}({\mathbf r})=\psi_{\downarrow}({\mathbf r})e^{ik_L x}$, where
$\big( \psi_{\uparrow}({\mathbf r}), \psi_{\downarrow}({\mathbf r}) \big)$ is a spinor wavefunction
on which Hamiltonian (\ref{app:scheme1_RF_2}) acts. The Hamiltonian then becomes:
\begin{equation}
\begin{split}
& \hat{H}_{RF,2} =\left[ \frac{\hbar^2 \hat{{\mathbf k}}^2}{2m} + V({\mathbf r}) 
- \omega_{r} \hat{L}_z +E_L \right] \check{1} \\
& \quad + \frac{\hbar^2 k_L \hat{k}_x}{m} \check{\sigma}_z
+ \frac{\Omega}{2} \check{\sigma}_x
+ \hbar \omega_{r} k_L y \check{\sigma}_z 
+ 
\begin{pmatrix}
0& 0\\
0& \hbar \omega_r - \delta
\end{pmatrix},
\end{split}
\label{app:scheme1_final}
\end{equation}
where $E_L=\hbar^2 k_{L}^2/2m$. We can drop $E_L \check{1}$ term 
by simply renormalizing the energy.

\section{M-scheme with rotating trap}
\label{app:scheme1_b}

The Hamiltonian $H_{\rm rot}^{\prime}$ describing M-scheme with rotating trap in the laboratory frame is:
\begin{equation}
\begin{split}
\hat{H}^{\prime} = & \left[ \frac{\hbar^2 \hat{{\mathbf k}}^2}{2m} + V \big( x(t),y(t),z \big) 
-\omega_r \hat{L}_z \right] \check{1} \\
& +\begin{pmatrix}
{\hbar(-\omega_z+\omega_q)}& 0& 0 \\
0& 0& 0 \\
0& 0& \hbar \omega_z 
\end{pmatrix} \\
& + \sqrt{2}\Omega \check{\sigma}_{3,x} \cos(2 k_L x + \Delta \omega_L t),
\end{split}
\label{eq:rotating_trap_Nature_1}
\end{equation}
where $x(t)$ is defined in (\ref{eq:xy_t}).
After transfering to the rotating frame ($\hat{U}(t)=\exp [i\omega_r t (\hat{L}_z + \hat{S}_z)/\hbar]$) 
and making the rotating wave approximation the Hamiltonian is: 
\begin{equation}
\begin{split}
\hat{H}_{RF}^{\prime} = & \left[ \frac{\hbar^2 \hat{{\mathbf k}}^2}{2m} + V({\mathbf r}) -\omega_r \hat{L}_z \right] 
\check{1} - \hbar \omega_r \check{\sigma}_{3,z} \\
& +\begin{pmatrix}
{3\delta/2 + \hbar \omega_q}& 0& 0 \\
0& \delta/2& 0 \\
0& 0& -\delta/2 
\end{pmatrix} \\
& + \frac{\Omega}{\sqrt{2}} \check{\sigma}_{3,x} \cos(2 k_L x^{\prime}(t) + \omega_r t) \\
& - \frac{\Omega}{\sqrt{2}} \check{\sigma}_{3,y} \sin(2 k_L x^{\prime}(t) + \omega_r t),
\end{split}
\label{eq:rotating_trap_Nature_2}
\end{equation}
where $x^{\prime}(t)=x \cos(\omega_r t) - y \sin(\omega_r t)$.
We may again neglect state $\lvert m_z=1 \rangle$ assuming $\omega_q>>$. To get the Hamiltonian
in a more familiar spin-orbit-coupling form we make the following transformation:
$\psi^{\prime}_{\uparrow}({\mathbf r})=\psi_{\uparrow}({\mathbf r})e^{-ik_L x^{\prime}(t)},
\psi^{\prime}_{\downarrow}({\mathbf r})=\psi_{\downarrow}({\mathbf r})e^{ik_L x^{\prime}(t)+i\omega_r t}$,
which gives:
\begin{equation}
\begin{split}
\hat{H}_{RF,2}^{\prime} =& \left[ \frac{\hbar^2 \hat{{\mathbf k}}^2}{2m} + V({\mathbf r}) 
- \omega_{r} \hat{L}_z +E_L \right] \check{1} \\
& +\frac{\hbar^2 k_L }{m}\hat{k}_x(t) \check{\sigma}_z
+ \frac{\Omega}{2} \check{\sigma}_x
+ \frac{\delta}{2} \check{\sigma}_z,
\end{split}
\label{eq:last_equation}
\end{equation}
where $\hat{k}_x(t)=\hat{k}_x \cos(\omega_r t) - \hat{k}_y \sin(\omega_r t)$.
We can drop $E_L \check{1}$ term by renormalizing the energy.

\section{Tripod scheme with rotating trap and spin-orbit lasers}
\label{app:scheme_2}

The original Hamiltonian for the tripod scheme (stationary system) is (see \cite{Stanescu2007}):
\begin{equation}
\hat{H}_0=  \frac{\hbar^2 \hat{{\mathbf k}}^2}{2m}\check{1} + \hat{V}({\mathbf r}) + \hat{H}_{a-l},
\end{equation}
where $\hat{V}({\mathbf r})=\sum_j V_j({\mathbf r})\lvert j \rangle \langle j \rvert$ is spin dependent
trapping potential, atom-laser interaction $\hat{H}_{a-l}=\Delta \lvert 0 \rangle \langle 0 \rvert - 
\big( \Omega_1 \lvert 0 \rangle \langle 1 \rvert + \Omega_2 \lvert 0 \rangle \langle 2 \rvert
+ \Omega_3 \lvert 0 \rangle \langle 3 \rvert + {\rm H.c.} \big)$ and $\check{1}$ is 4 $\times$ 4 unit matrix.
$\Delta$ is detuning from resonance and $\Omega_{1,2,3}$ are Rabi frequencies:
$\Omega_1({\mathbf r})=\Omega \sin \theta \cos(m v_a x) e^{im v_b y}$,
$\Omega_2({\mathbf r})=\Omega \sin \theta \sin(m v_a x) e^{im v_b y}$,
$\Omega_1({\mathbf r})=\Omega \cos \theta$, where $\Omega$, $\theta$, $v_a$ and $v_b$
are constants (see \cite{Stanescu2007} for details). If we start rotating spin-orbit lasers
in the laboratory, atom-laser interaction part of the Hamiltonian becomes 
$e^{-i\omega_r t (\hat{L}_z+\hat{S}_z)/\hbar}\hat{H}_{a-l}e^{i\omega_r t(\hat{L}_z+\hat{S}_z)/\hbar}$.
If the trap rotates, trapping potential becomes $e^{-i\omega_r t (\hat{L}_z+\hat{S}_z)/\hbar}
\hat{V} e^{i\omega_r t (\hat{L}_z+\hat{S}_z)/\hbar}$. 
Therefore, we can write the Hamiltonian of the rotating system as: 
\begin{equation}
\hat{H}_{\rm rot}=e^{-i \omega_r t (\hat{L}_z+\hat{S}_z)/\hbar} \hat{H}_0 e^{i \omega_r t (\hat{L}_z+\hat{S}_z)/\hbar}.
\end{equation}
The Hamiltonian in the rotating frame is then: $\hat{H}_{RF}=\hat{H}_0-\omega_r(\hat{L}_z+\hat{S}_z)$.  
Since $\hat{H}_{RF}$ is time-independent we can use exactly the same procedure for getting 
the effective spin-orbit coupling described in \cite{Stanescu2007,Ruseckas2005}, i.e. we project the
Hamiltonian to the dark states subspace. Here we assume that three degenerate hyperfine groundstates
are part of F=1 manifold (for example the ground state of $^{87}$Rb) and that they are eigenstates of $\hat{S}_z$. 
This gives us the precise form of $\hat{S}_z$ operator. 
As in \cite{Stanescu2007} we take $V_1=V_2=w({\mathbf r})$ and $V_3=w({\mathbf r})+\delta$.
After projecting to dark states we get:
\begin{equation}
\begin{split}
\hat{H}_{RF,2} &=\left[ \frac{\hat{{\mathbf p}}^2}{2m} + w({\mathbf r}) - \hat{L}_z \right]\check{1} 
-v_0 \hat{p}_x \check{\sigma}_y - v_1 \hat{p}_y \check{\sigma}_z  \\
& + \delta_0 \check{\sigma}_z + m \hbar \omega_r (v_1 x \check{\sigma}_z - v_0 y \check{\sigma}_y) \\
&- \hbar \omega_r
\begin{pmatrix}
\sin^2 \phi & \sin \phi \cos \phi \cos \theta \\
\sin \phi \cos \phi \cos \theta & \cos^2 \theta \cos^2 \phi - \sin^2 \theta
\end{pmatrix},
\end{split}
\label{app:tripod_final}
\end{equation}
where $\delta_0=\sin^2 \theta \big \lbrace \delta - \big[ \big( \frac{v_0}{\cos \theta} \big)^2 + 
\big( \frac{v_1}{\sin^2 (\theta/2)} \big)^2 \big]/2 \big \rbrace /2$. 
 
\end{appendix}

\bibliography{references}

\end{document}